\documentclass[reprint,a4paper]{article}

\usepackage[numbers,sort&compress]{natbib}
\usepackage[affil-it]{authblk}
\usepackage{amsmath,amsfonts,amssymb}
\usepackage{algorithm,algorithmic}
\usepackage{upgreek,cancel}
\usepackage{graphicx}
\usepackage[dvipsnames]{xcolor}
\usepackage{float}
\usepackage{tikz}
\usepackage{booktabs}
\usepackage{pbox}
\usepackage{caption,subcaption}
\usepackage{setspace,verbatim,comment}
\usepackage[colorlinks]{hyperref}
\hypersetup{
    colorlinks=true,
    linkcolor=blue,
    filecolor=magenta,
    urlcolor=[rgb]{0.00,0.00,0.50},
    citecolor=red,
    }
\usepackage[nameinlink]{cleveref}
\usepackage[title]{appendix}
\usepackage[top=2cm, bottom=2cm, left=3cm, right=3cm]{geometry}

\newcommand{\ii}{\ensuremath{\mathrm{i}}}
\definecolor{darkgreen}{rgb}{0,0.35,0}

\DeclareMathOperator{\sech}{sech}
\DeclareMathOperator{\sn}{sn}

\def\tsc#1{\csdef{#1}{\textsc{\lowercase{#1}}\xspace}}
\tsc{WGM}
\tsc{QE}
\providecommand{\keywords}[1]
{
\small
\textbf{\textit{Keywords---}} #1
}
\newcommand{\CECs}{Centro de Estudios Cient\'{\i}ficos (CECs), Casilla 1469, Valdivia, Chile}
\newcommand{\USS}{Universidad San Sebasti\'{a}n, sede Valdivia, General Lagos 1163, Valdivia 5110693, Chile}
\newcommand{\UdeC}{Universidad de Concepci\'{o}n, Concepci\'{o}n, Chile}
\newcommand{\UACh}{Instituto de Ciencias F\'isicas y Matem\'aticas, Universidad Austral  de Chile, Casilla 567, Valdivia, Chile}
\newcommand{\UCharles}{IPNP - Faculty of Mathematics and Physics, Charles University, V Hole\v{s}ovi\v{c}k\'ach 2, 18000 Prague 8, Czech Republic}

\begin{document}
\let\WriteBookmarks\relax
\def\floatpagepagefraction{1}
\def\textpagefraction{.001}

% Main title of the paper
\title{BPS baryonic tubes in chiral perturbation theory minimally coupled with electrodynamics and a novel crossover}

\author[1,2]{Fabrizio Canfora\thanks{fabrizio.canfora@uss.cl}}
\author[3]{Mart\'{\i}n Garrido\thanks{margarrido2021@udec.cl}}
\author[4,5]{Pablo Pais\thanks{pais@ipnp.troja.mff.cuni.cz}}
\date{}
\affil[1]{\CECs} \affil[2]{\USS} \affil[3]{\UdeC} \affil[4]{\UACh} \affil[5]{\UCharles}

\maketitle

% Here goes the abstract
\begin{abstract}
We derive a novel first-order BPS system (whose solutions are automatically solutions of the second-order field equations) in chiral perturbation theory (CPT) in $(3+1)$ dimensions. These BPS configurations describe baryonic tubes carrying a chiral current along the tubes themselves. Such baryonic tubes can be either homogeneous or modulated along the $z$-axis. We analyze the leading term in CPT, which is the $SU(2)$ non-linear sigma model minimally coupled to electrodynamics. These exact solutions are qualitatively different from the available analytic description of baryonic tubes since, due to their BPS nature, the positions of the energy-density peaks can be arbitrary. In contrast (in the case of the known solutions in the literature), the energy-density peaks' positions are fixed. We compare these two types of baryonic configurations; the relevant parameter that determines which one is energetically favored is the baryonic charge density: for low densities, the old solutions are favorable, while for high densities these novel BPS tubes are energetically favored. The inclusion of the electromagnetic field (together with the corresponding back-reaction on the baryonic tubes) can also be treated analytically. We discuss the differences in the electromagnetic field distribution between the two regions (low and high baryonic density). The analysis of potential terms for the chiral field is detailed in Appendix \ref{app:potential_term}.
\end{abstract}


\newpage

\section{Introduction}\label{Intro}

The analysis of finite-density effects in the low-energy limit of QCD (especially when the interaction with strong electromagnetic fields is included) is an extremely important but difficult topic due to the limitations of both perturbative and Lattice QCD (LQCD) methods (see \cite{R0,R11,R2,sign1,sign2,sign3m,Astrakhantsev:2021uii,Brandt:2023dir,HIC,QGBook} and references therein). One of the reasons is that, at low energies, non-perturbative solitonic configurations have a dominant role in the study
of the phase diagram (see~\cite{R11,MantonSutcliffe,ShifmanBook} for details) so that, in these cases, QCD perturbation theory is not effective.
Such configurations possess non-trivial topological charge(s), which make it impossible for them to decay into the trivial vacuum. A very intriguing finite-density effect in the low-energy phase diagram of QCD is the formation of baryonic condensates, which, unlike what happens in free space,
are not homogeneous. Results such as~\cite%
{Brauner:2016pko,KawaguchiC100,Brauner:2019aid,Gronli:2022cri,Higaki2022,Caslbuoni2004,pasta1,pasta2,pasta3,pasta4,Dorso:2018lkv,Dorso:2020zhk,Lopez:2020zne,Pelicer:2022ncw,Nandi:2017aqq,Yakovlev:2015vma}
strongly suggest that non-homogeneous condensates are a typical feature of the low-energy phase diagram of QCD. It is worth emphasizing that this also happens in BCS superconductors~\cite{LOFF1,LOFF2}. The non-homogeneous
baryonic condensates that appear at finite baryon density exhibit beautiful regular shapes. Both numerical simulations and indirect astrophysical observations are consistent with the formation of regular arrays of baryonic tubes and layers (known as \emph{nuclear spaghetti} and \emph{nuclear
lasagna phases}).

Numerical simulations of such configurations are already extremely complicated even when the electromagnetic field is turned off; needless to say, including it on top of nuclear interactions makes matters much worse.
Consequently, new effective theoretical techniques can complement numerical analysis and simulations; the development of such techniques is the main task of the present manuscript. As it will be discussed in the following sections, the importance of the analytical tools which will be introduced in the present manuscript goes far beyond the mere support of the
numerical results and simulations. Indeed, we will be able to identify
different kinds of baryonic tubes as well as a novel cross-over phenomenon
in which, depending on the baryon density, one type of tube is favored over the other. Such a phenomenon could not have been discovered with numerical techniques. The behavior of the electromagnetic field changes a lot depending on which kind of tubes is energetically favored as well.

The starting point of our analysis will be Chiral Perturbation Theory (CPT)
\cite{CPT1,CPT2,CPT3,1m,2m,3m,4m}, which represents the low-energy limit of
QCD in $(3+1)$ dimensions. The leading term of CPT is the $(3+1)$%
-dimensional $SU(2)$ gauged non-linear sigma model (gauged NLSM henceforth)
minimally coupled to Maxwell theory. Besides its intrinsic interest in
relation to the low-energy limit of QCD, such a theory is also very
interesting in itself (for a detailed discussion, we refer the reader to
\cite{MantonSutcliffe,BaMa}). In the pioneering references \cite%
{Skyrme1,Skyrme2,Skyrme3,Skyrme4}, it was shown that the topological charge
of this theory corresponds to the baryonic charge. The inclusion of a
potential term will also be explored. Consequently, determining the physical
requirements that lead to the formation of non-homogeneous baryonic
condensates is a fundamental task in the analysis of dense nuclear/quark
matter \cite{R0,R2,ex4d6,ex4d4}.

In $(1+1)$-dimensions, many examples are available in which, at finite
density, ordered arrays of kinks dominate a finite region of the phase
diagram~\cite%
{Dunne111,Basar105012,Schnetz:2005ih,Thies2018,Thies:2019ejd,Thies2020,Thies096006,Thies:2023dyh,Adhikari:2017ydi}%
. A very nice example of an inhomogeneous baryonic condensate is the Chiral
Soliton Lattice (CSL)~\cite{Brauner:2016pko,ex4d5,ex4d1,ex4d3}; however, in
this case, the pionic field only depends on one spatial coordinate. At first
glance, it is extremely challenging to obtain similar analytic results when
the hadronic profiles depend on more than one coordinate (perhaps this
impression should not be too surprising, given that CPT is the prototype of
non-integrable theories). It is a common belief that the standard approaches
to deal with non-perturbative configurations (which work very well, for
instance, in the case of Yang-Mills-Higgs and Abelian-Higgs theories in the
analysis of instantons, monopoles, and vortices) cannot be applied
straightforwardly in CPT. The main circumstance behind such a belief is
that, in CPT minimally coupled with Maxwell gauge theory, (until very
recently) there was no obvious BPS bound to saturate. As it is well known, in
the theory of non-Abelian monopoles, instantons, critical vortices in
superconductors and so on, the availability of a first-order BPS system
plays a fundamental role. In fact, in the case of gauged CPT, if one puts
on the right-hand side of the sought BPS bound(s) the \textquotedblleft
obvious\textquotedblright\ topological charges (such as the baryonic charge
and the magnetic flux), the BPS formalism is not effective.

For these reasons, in CPT it has been necessary to introduce (see~\cite{canfora2,56,56b,SupercCrystals,58,58b,ACZ,canfora10,Fab1,gaugsk,Canfora:2018clt,lastEPJC,crystal1})
different analytic techniques which are not directly related to the BPS
formalism. Such analytic approaches yield several analytic configurations
describing inhomogeneous baryonic condensates with the shapes of baryonic
layers and tubes together with their corresponding electromagnetic fields in
a self-consistent way (the properties of these analytic solutions are in
good agreement with the results in \cite%
{pasta1,pasta2,pasta3,pasta4,Lopez2018,Dorso:2020zhk,Lopez:2020zne,Pelicer:2022ncw,Nandi:2017aqq,Yakovlev:2015vma}%
). However, in this approach (due to the lack of a saturable BPS bound)
stability must be analysed separately (see the detailed numerical discussion
in \cite{stability}). Consequently, we are interested in the following
important question:

\textit{Can we find novel BPS bounds in CPT in terms of different
topological charges which can actually be saturated?}

The importance of this question is two-fold. Not only does the stability
analysis become much more transparent, but also many powerful tools from the
theory of BPS states can be ``imported''~\cite%
{R11,MantonSutcliffe,ShifmanBook,Shuryak}.

Needless to say, the most natural candidate to appear on the right-hand side
of the BPS bound (which, in the present case, is the baryonic charge) is not
working. Moreover, in the case of CPT, the numerical configurations which
have been constructed so far are above the obvious bound in terms of the
baryonic charge~\cite{R11,MantonSutcliffe,ShifmanBook,Shuryak}.
Nevertheless, quite remarkably, in~\cite{BPS1,BPS2,BPS3,BPS4}, the first examples
of non-obvious BPS bounds have been found which can actually be saturated
and which possess the sought property~\cite%
{Alonso-Izquierdo2023,Alonso-Izquierdo2026} that the saturation of the bound
implies the second-order field equations. In the present manuscript, we will
find the first example of a BPS system and the corresponding BPS bound
suitable to deal with baryonic tubes. The configurations which satisfy this
first order BPS system are baryonic tubes which, however, are different\footnotemark[1] from
the ones already known in the literature (constructed in \cite{56} and in
\cite{OldSolutions}). This discovery disclose a novel cross-over
phenomenon: for low baryonic density the "old tubes" are energetically
favored while for high baryonic density the present BPS tubes are favored. \footnotetext[1]{
From the technical viewpoint, it is also important to emphasize the scarcity
(in the low energy limit of QCD) of multi-baryonic BPS solutions with
arbitrarily chosen energy-peak positions, similar to multi-instantons in
Yang-Mills theory (compare with \cite{BPS1} and \cite{OldSolutions}).}

This paper is organized as follows: in Section 2, we introduce the leading
term of chiral perturbation theory minimally coupled to electrodynamics. In
Section 3, we derive the novel BPS bound for baryonic tubes, analyze the
complex structure of the field equations, and compare these BPS states with
rigid soliton crystals. In Section 4, we extend these configurations to the
gauged non-linear sigma model. Finally, Section 5 presents our conclusions
and perspectives. The generalization including potential terms is detailed
in Appendix \ref{app:potential_term}.

\section{The leading term in Chiral Perturbation Theory minimally coupled with electrodynamics}

According to Chiral Perturbation Theory ($\chi $PT)~\cite{1m,2m,3m,4m}), at
leading order the low-energy QCD action\footnotemark[2] reads \footnotetext[2]{
We have chosen natural units; $c=1$, $\hslash =1$.}
\begin{equation}
S=\int \frac{d^{4}x}{4}\sqrt{-g}\left\{ \widehat{K}\ \mathrm{Tr}\left[
L^{\mu }L_{\mu }\right] -\frac{1}{e^{2}}F_{\mu \nu }F^{\mu \nu }\right\} ,
\label{action0}
\end{equation}%
where $e$ is the electric charge and $\widehat{K}$ is the coupling constant
of the gauged NLSM. The constant $\widehat{K}$ is related to the pion decay
constant $f_{\pi }$ as follows (see \cite{Skyrme4}):
\begin{equation}
\widehat{K}=\frac{\left( f_{\pi }\right) ^{2}}{4}\ ,\ \ \ f_{\pi }\approx
130\ \ MeV\ .  \label{action0.1}
\end{equation}

The action of the $SU(2)$ gauged NLSM system is
\begin{equation}
S=\int d^{4}x\sqrt{-g}\left[\frac{K}{4} \,\ \mathrm{Tr}\left(
L^{\mu}L_{\mu}\right) -\frac{1}{4}F_{\mu\nu }F^{\mu\nu} \right] ,\quad K=e^2%
\widehat{K}   \label{NLSMaction}
\end{equation}
where $K$ is the "rescaled" pion decay constant, and
\begin{align}
L_{\mu}:= & U^{-1}D_{\mu}U=L_{\mu}^{j}t_{j},\ \ D_{\mu}U=\nabla_{\mu
}U+A_{\mu}\left[ t_{3},U\right] \;, \\
\ \ F_{\mu\nu}=&\partial_{\mu}A_{\nu }-\partial_{\nu}A_{\mu}, \ \
t_{j}=i\sigma_{j} \;,
\end{align}
with $g$ being the metric determinant, $A_\mu$ is the gauge potential, and $%
\sigma_i$ are the Pauli matrices. The field equations are
\begin{align}
D_{\mu} L^{\mu}& =0\ , \\
\nabla_{\mu}F^{\mu\nu} & =J_{U(1)}^{\nu}\;,
\end{align}
where the $U(1)$ current is
\begin{equation}
J_{U(1)}^{\nu}=\frac{K}{2} Tr[( U^{-1}t_{3} U-t_{3})L^{\nu}]\;.
\end{equation}
The baryonic charge is given by
\begin{equation}
B=\frac{1}{24\pi^{2}}\int_{t = const}\rho_{B},
\end{equation}
where
\begin{align}
\rho_B&=\rho_B^{NLSM}+\rho_B^{Maxwell},  \notag \\
&=\epsilon^{ijk} \mathrm{Tr}\bigg[\big( U^{-1}\partial_{i}U\big)\big( %
U^{-1}\partial_{j}U\big)\big( U^{-1}\partial_{k}U\big)-\partial_i\Big[3A_jt_3%
\big(U^{-1}\partial_kU+(\partial_k U)U^{-1}\big)\Big]\bigg],
\end{align}
is the baryonic charge density. The stress-energy tensor is
\begin{equation}
T_{\mu\nu}=-\frac{K}{2}\left[ L_\mu L_\nu - \frac{1}{2}g_{\mu\nu}L^{%
\alpha}L_{\alpha} \right]+\bar{T}_{\mu\nu},
\end{equation}
where $\bar{T}_{\mu\nu}=F_{\mu\alpha}F_{\nu}^{\,\alpha}-\frac{1}{4}%
F_{\alpha\beta}F^{\alpha\beta}g_{\mu\nu}$ is the electromagnetic
stress-energy tensor.
\subsection{Exponential Parametrization}
\label{Param} In what follows, we will adopt the exponential parametrization
of the elements of $SU(2)$. Without loss of generality, any element of the $%
SU(2)$ group can be written as:
\begin{align}
U& =\cos \alpha \mathbf{I}+\sin \alpha n^{j}t_{j}\ , \\
\vec{n}& =\left( \sin \Theta \cos \Phi ,\sin \Theta \sin \Phi ,\cos \Theta
\right)
\end{align}%
In terms of the degrees of freedom $\alpha $, $\Theta $ and $\Phi $ above
exponential parametrization, the $U(1)$ covariant derivative $D_{\mu
}U=\nabla _{\mu }U+A_{\mu }\left[ t_{3},U\right] $ reads:%
\begin{align}
D_{\mu }\alpha & =\nabla _{\mu }\alpha  \\
D_{\mu }\Theta & =\nabla _{\mu }\Theta  \\
D_{\mu }\Phi & =\nabla _{\mu }\Phi -2A_{\mu }
\end{align}%
The gauged NLSM action is then defined
\begin{equation}
S=-\frac{1}{2}\int d^{4}x\sqrt{-g}\bigg\{K\big(\nabla _{\mu }\alpha \nabla
^{\mu }\alpha +\sin ^{2}\alpha \nabla _{\mu }\Theta \nabla ^{\mu }\Theta
+\sin ^{2}\alpha \sin ^{2}\Theta D_{\mu }\Phi D^{\mu }\Phi \big)+\frac{1}{2}%
F_{\mu \nu }F^{\mu \nu }\bigg\}.
\end{equation}%
The field equation corresponding to the variation with respect to $\alpha $,
$\Theta $ and $\Phi $ read
\begin{align}
& -\square \alpha +\sin \alpha \cos \alpha \left( \partial _{\mu }\Theta
\partial ^{\mu }\Theta +\sin ^{2}\Theta D_{\mu }\Phi D^{\mu }\Phi \right) =0,
\label{equalpha} \\
& -\sin ^{2}\alpha \square \Theta -2\sin \alpha \cos \alpha \partial _{\mu
}\alpha \partial ^{\mu }\Theta +\sin ^{2}\alpha \sin \Theta \cos \Theta
D_{\mu }\Phi D^{\mu }\Phi =0, \\
& -\sin ^{2}\alpha \sin ^{2}\Theta \partial _{\mu }D^{\mu }\Phi -2\sin
\alpha \cos \alpha \sin ^{2}\Theta \partial _{\mu }\alpha D^{\mu }\Phi
-2\sin ^{2}\alpha \sin \Theta \cos \Theta \partial _{\mu }\Theta D^{\mu
}\Phi =0,
\end{align}%
while the Maxwell equations (the field equation corresponding to the
variation with respect to $A_{\mu }$) is
\begin{equation}
    \nabla_{\mu}F^{\mu\nu}=-2K\sin^{2}\alpha\sin^2\Theta\,(\nabla^\nu\Phi-2A^\nu).
\end{equation}
In terms of these degrees of freedom, the baryonic charge
density reads
\begin{equation}
\rho _{B}=12\sin ^{2}\alpha \sin \Theta \,d\alpha \wedge d\Theta \wedge
d\Phi +12\,d\left[ \boldsymbol{A}\wedge \left( \cos \Theta \,d\alpha -\frac{1%
}{2}\sin (2\alpha )\sin \Theta \,d\Theta \right) \right] \ ,
\end{equation}
where the second term ensures it is gauge invariant.

Indeed, calling $\omega=\cos \Theta \,d\alpha -\frac{1
}{2}\sin (2\alpha )\sin \Theta \,d\Theta $, we can compute its exterior derivative
\begin{align}
d\omega&=d\big(\cos \Theta \,d\alpha\big) - \frac{1
}{2}\, d \big(\sin (2\alpha )\sin \Theta \,d\Theta\big), \nonumber\\
&=-\sin\Theta d\Theta\wedge d\alpha - \cos(2\alpha)\sin\Theta d\alpha\wedge d\Theta, \nonumber \\
&=\big( 1-\cos(2\alpha) \big)\sin\Theta d\alpha\wedge d\Theta, \nonumber \\
&=2\sin^2\alpha\sin\Theta d\alpha\wedge d\Theta.
\end{align}
Then, under the gauge transform
\begin{align}
 A_{\mu } &\rightarrow A_{\mu }+\nabla _{\mu }\Lambda,  \\
\Phi &\rightarrow \Phi +2\Lambda ,
\end{align}
the variation of $\rho_B$ is
\begin{align}
\Delta\rho_B&=24\sin ^{2}\alpha \sin \Theta \,d\alpha \wedge d\Theta \wedge
d\Lambda - 12d\Lambda\wedge d\omega,\nonumber \\
&=0.
\end{align}
Thus, in the case of $A_{\mu }=0$, in order for a configuration to have
non-trivial baryonic charge it will be crucial that the fields $\alpha $, $%
\Theta $ and $\Phi $ are three independent functions. On the other hand, one
has to require that the spatial integral of $\rho _{B}$ must be a
non-vanishing integer as well.

\section{BPS Method and Chiral Perturbation Theory}
The leading term in CPT is the non-linear sigma model (NLSM) with the $SU(2)
$ isospin symmetry (in the next sections we will discuss the role of the Maxwell gauge field, while in the Appendix \ref{app:potential_term} we will
discuss the role of potential-like terms). A very relevant issue (which
arises at finite baryon density) concerns the emergence of baryonic
inhomogeneous condensates: this issue plays a central role in the present
manuscript. As emphasised in previous sections, for several years it was
assumed that no useful BPS bound existed in CPT. In fact, quite recently,
in~\cite{BPS1,BPS2,BPS3,BPS4} the first examples of non-obvious BPS bounds
in CPT have been constructed. The idea is to insist that a BPS bound should
exist, but without assuming that the right-hand side of the bound itself is
given in terms of the obvious topological charge: indeed, the field
equations themselves show the way.
\subsection{Ansatz and BPS bound}

Since we aim to describe non-homogeneous condensates at finite density,
finite-volume effects are crucial. We can take account of them by
considering the flat metric
\begin{equation}
ds^{2}=-dt^{2}+L^{2}(dx^{2}+dy^{2})+L_{z}^{2}dz^{2},
\end{equation}%
where the scale factor $L_{z}$ is introduced to keep track of the $z$%
-direction; throughout this work, we set $L_{z}=L$. The volume of the box in
which the topological solitons are living is $4\pi ^{3}L^{2}L_{z}$. The
adimensional coordinates are
\begin{equation}
0\leq x\leq 2\pi ,\quad 0\leq y\leq \pi ,\quad 0\leq z\leq 2\pi .
\end{equation}%
This setup will also allow us to compare our solutions to those in \cite{OldSolutions}.
As it will be clear in the following analysis, the tubes extend along the
direction defined by $z$, while the dynamics reduces to a two-dimensional
description in the coordinates $x$ and $y$.

We begin by considering the following ansatz
\begin{equation}
\alpha =\alpha (x,y),\quad \Theta =\Theta (x,y),\quad \Phi =p\left( \frac{t}{%
L_{z}}-z\right) ,\quad A_{\mu }=(u(x,y),0,0,-L_{z}u(x,y))
\end{equation}%
which is a generalization of the ansatz proposed in \cite{OldSolutions} that
preserves relevant properties such as the periodic time dependence, which
allows one to evade Derrick's theorem while still maintaining a stationary
stress-energy tensor and electromagnetic field. Note in particular that both
the function $\Phi $ and the ansatz for $A_{\mu }$ are the same as in \cite%
{OldSolutions} (this choice allows to compare easily the energies of these
two configurations), while $\alpha $ and $\Theta $ are more general as they
can depend on both spatial coordinates (in \cite{OldSolutions} $\alpha
=\alpha (x)$ and $\Theta =\Theta (y)$). One can observe that, with the above
ansatz the energy-momentum tensor does depend neither on $t$ nor on $z$. On
the other hand, one can also construct analytically BPS baryonic tubes which
are modulated along the $z$-axis summing to $\Phi $ an arbitrary function of
the light-like variable $\left( \frac{t}{L_{z}}-z\right) $. We will not
discuss this case here, however we observe that these modulations along the $%
z$-axis are associated to a chiral superfluid current proportional to $%
\nabla _{\mu }\Phi $ (see the analysis in \cite{SupercCrystals}). Another relevant property of this ansatz (in which $\Phi $ is a
linear function of the light-like coordinate $\left( \frac{t}{L_{z}}%
-z\right) $) is the fact that the relevant observables (such as the
energy-momentum tensor and the baryon density) only depend on $x$ and $y$.
Nevertheless, the explicit time-dependence of the chiral field is already providing a relevant information: since the configuration is not static, we should not expect to find the correct BPS bound analyzing the energy density (as this approach only works for static configurations) but rather by analyzing the action, which plays the role of free energy.

The above ansatz satisfies the following useful relations
\begin{align}
A_{\mu }A^{\mu }=0,\quad \partial _{\mu }\Phi \partial ^{\mu }\Phi =&
0,\quad A_{\mu }\partial ^{\mu }\Phi =0\ , \\
\partial _{\mu }\alpha \partial ^{\mu }\Phi =\partial _{\mu }\Theta \partial
^{\mu }\Phi =0,& \quad A_{\mu }\partial ^{\mu }\alpha =A_{\mu }\partial
^{\mu }\Theta =0\ .
\end{align}
Using the above ansatz the complete set of field equations of the gauged
NLSM-Maxwell theory reduce to the following coupled system of three PDEs:
\begin{align}
-\Delta \alpha +\sin \alpha \cos \alpha (\boldsymbol{\nabla }\Theta )^{2}=&
0, \\
-\sin ^{2}\alpha \Delta \Theta -2\sin \alpha \cos \alpha (\boldsymbol{\nabla
}\alpha \cdot \boldsymbol{\nabla }\Theta )=& 0,\\
\boldsymbol{\Delta} \Psi +V\Psi =&0 \label{maxwell},
\end{align}%
where $V=4L_{z}^{2}K\sin ^{2}(\alpha )\sin
^{2}(\Theta )$ and we have defined the auxiliary field $\Psi =\frac{2L_{z}}{p}u-1$.

\subsection{The case of vanishing electromagnetic field}

Here we will consider the case of vanishing electromagnetic field (the
inclusion of a non-trivial $F_{\mu \nu }$ will be discussed in the following
sections\footnotemark[3])\footnotetext[3]{Regardless, our ansatz choice for $A_\mu$ correspond to a pure-radiation configuration $F_{\mu \nu }F^{\mu \nu }=0$, ensuring that our present construction is not spoiled in the case of non-vanishing electromagnetic field.}. The effective Lagrangian (namely, the Lagrangian evaluated on the
ansatz $\alpha (x,y)$, $\Theta (x,y)$, $\Phi =p\left( \frac{t}{L_{z}}%
-z\right) $) can be rewritten as
\begin{equation}
\mathcal{L}=\frac{K}{2L^{2}}\big[(\partial _{x}\alpha \pm \sin \alpha
\partial _{y}\Theta )^{2}+(\partial _{y}\alpha \pm \sin \alpha \partial
_{x}\Theta )^{2}\big]\pm \frac{K}{L^{2}}\sin \alpha (\partial _{y}\alpha \partial _{x}\Theta
-\partial _{x}\alpha \partial _{y}\Theta ),
\end{equation}%
where the last term is a total derivative
\begin{equation}\label{BPSden1}
\pm\sin \alpha (\partial _{y}\alpha \partial _{x}\Theta
-\partial _{x}\alpha \partial _{y}\Theta )dx\wedge dy=\pm\sin\alpha d\Theta\wedge d\alpha=\pm d(\cos\alpha d\Theta).
\end{equation}

A direct computation shows that all the configurations satisfying the
following first-order BPS equations
\begin{align}
\partial _{x}\alpha \pm \sin \alpha \,\partial _{y}\Theta & =0\;,
\label{BPS1} \\
\partial _{y}\alpha \mp \sin \alpha \,\partial _{x}\Theta & =0\;,
\label{BPS2}
\end{align}%
automatically satisfy the second-order field equations. As it has been
already emphasized, it is natural that, in the present case in which the
chiral field depends explicitly on time, there is no saturable BPS bound on
the energy while the correct BPS bound is associated to the effective
two-dimensional action.

The total BPS charge can be evaluated via Stokes' theorem over the domain $%
R\setminus \bigcup D_{j}$ (where $D_{j}$ are small disks around the
positions of the $j$-th tube, as it will be explained here below), taking
the limit $\epsilon _{j}\rightarrow 0$ around each defect one obtains
\begin{equation}
Q_{\text{BPS}}=\oint_{\partial R}\cos \alpha \,d\Theta +2\pi n
\end{equation}%
where the boundary integral captures the regular background contribution and
the $+2\pi n$ term accounts for the sum over the internal defects, as will
be seen in section \ref{BPSsol}. From the BPS decomposition for the
effective action one obtains
\begin{equation}
\int_{R}\mathcal{L}\,dx\,dy\ \geq  K|Q_{BPS}|\ .
\end{equation}

\subsection{Stress-Energy tensor}
\label{section_stress-energy_tensor}
In this subsection the electromagnetic
field is turned on again.

First of all, it is convenient to simplify the expression of the energy per
unit of length $\frac{E}{2\pi L_{z}}$ using the above BPS decomposition: one
can start evaluating the following difference
\begin{equation}
T_{00}-\mathcal{L}=\frac{1}{L^{2}}(\boldsymbol{\nabla }u)^{2}+\frac{K}{%
L_{z}^{2}}\sin ^{2}\alpha \sin ^{2}\Theta \left( p-2L_{z}u\right) ^{2}.
\label{EdenXterm}
\end{equation}%
which guarantees that the energy per unit of length is bounded from below.
Moreover, after integration, one gets
\begin{equation}
\frac{E}{2\pi L_{z}}=K|Q_{BPS}|+K\int_{R}\sin ^{2}\alpha \sin ^{2}\Theta
\left( p-2L_{z}u\right) ^{2}\,dx\,dy+\int_{R}(\boldsymbol{\nabla }%
u)^{2}\,dx\,dy.
\end{equation}

As far as the stress-energy tensor is concerned, one obtains the following
expressions for the non-vanishing components:
\begin{align}
T_{00}=& \frac{K}{2L^{2}}\left[ (\boldsymbol{\nabla }\alpha )^{2}+\sin
^{2}\alpha (\boldsymbol{\nabla }\Theta )^{2}\right] , \\
T_{xx}=& \frac{K}{2}\big((\partial _{x}\alpha )^{2}-(\partial _{y}\alpha
)^{2}+\sin ^{2}\alpha (\partial _{x}\Theta )^{2}-\sin ^{2}\alpha (\partial
_{y}\Theta )^{2}\big), \\
T_{yy}=& \frac{K}{2}\big(-(\partial _{x}\alpha )^{2}+(\partial _{y}\alpha
)^{2}-\sin ^{2}\alpha (\partial _{x}\Theta )^{2}+\sin ^{2}\alpha (\partial
_{y}\Theta )^{2}\big), \\
T_{xy}=& \;\partial _{x}\alpha \partial _{y}\alpha +\sin ^{2}\alpha \partial
_{x}\Theta \partial _{y}\Theta , \\
T_{zz}=& -\frac{K}{2}\left[ (\boldsymbol{\nabla }\alpha )^{2}+\sin
^{2}\alpha \,(\boldsymbol{\nabla }\Theta )^{2}\right] ,
\end{align}%
while the components of the electromagnetic stress-energy tensor are
\begin{align}
\bar{T}_{00}=& \frac{1}{L^{2}}(\boldsymbol{\nabla }u)^{2}+%
\frac{K}{L_{z}^{2}}\sin ^{2}\alpha \sin ^{2}\Theta \left( p-2L_{z}u\right)
^{2}, \\
\bar{T}_{0z}=& -\frac{L_{z}}{L^{2}}(\boldsymbol{\nabla }u)^{2}-\frac{K}{L_{z}%
}\sin ^{2}\alpha \sin ^{2}\Theta \,(p-2L_{z}u)^{2}, \\
\bar{T}_{zz}=& (\boldsymbol{\nabla }u)^{2}+K\sin ^{2}\alpha \sin ^{2}\Theta
\,(p-2L_{z}u)^{2}\;,
\end{align}%
On the other hand, in the case of configurations satisfying the first order
BPS equations, the energy-momentum tensor simplifies considerably. If one
defines $\rho _{BPS}$ as (choosing the upper sign in \ref{BPSden1})
\begin{equation}
\rho _{BPS}\equiv \sin \alpha (\partial _{y}\alpha \partial _{x}\Theta
-\partial _{x}\alpha \partial _{y}\Theta )= \sin ^{2}\alpha (\boldsymbol{%
\nabla }\Theta )^{2},
\end{equation}
the total stress-energy tensor takes the simple form
\begin{equation}
\begin{pmatrix}
\frac{K}{2L^{2}}\rho _{BPS}+\bar{T}_{00} & 0 & 0 & \bar{T}_{0z} \\
0 & 0 & 0 & 0 \\
0 & 0 & 0 & 0 \\
\bar{T}_{0z} & 0 & 0 & -\frac{K}{2}\rho _{BPS}+\bar{T}_{zz}%
\end{pmatrix}%
\ .
\end{equation}%
Consequently, the transverse pressures vanish ($T_{xx}=P_{x}=0$ and $%
T_{yy}=P_{y}=0$), confirming the absence of net inter-tubes forces and
ensuring the static equilibrium of multi-tube configurations at arbitrary
transverse positions. The non-zero entries of $T_{\mu \nu }$ are restricted
to the longitudinal-temporal sector, where $T_{00}$ represents the total
energy density, $\bar{T}_{zz}$ the longitudinal tension, and $\bar{T}_{0z}$
a persistent Poynting momentum flux along the tube axis.

\section{BPS solitons with non-trivial baryonic charge}
In this section, we will construct analytic solutions to our BPS equations
with non-trivial baryonic charge and explore their physical properties. The
following boundary condition will be imposed for the Maxwell potential
\begin{equation}
u(x,y)\big|_{\partial R}=0,
\end{equation}%
ensuring that the baryonic charge comes exclusively from the NLSM sector
\begin{equation}
B=-\frac{p}{\pi }\int_{R}\sin ^{2}\alpha \sin \Theta d\alpha \wedge d\Theta
=-\frac{p}{2\pi }\oint_{\partial R}(\alpha -\frac{1}{2}\sin 2\alpha )\sin
\Theta d\Theta ,
\end{equation}%
while the contribution from the Maxwell term vanishes, and keeping our
analysis consistent. Of course, this is not the only option but we have
chosen it in order to compare the present novel configurations with the ones
which have been already analyzed in the literature in \cite{OldSolutions}, as it would be
difficult to compare the energies of two configurations with different
behaviors of the electromagnetic field at the boundaries.

\subsection{BPS equations and complex structure}

The fields $\alpha $ and $\Theta $ satisfy the BPS equations (which, as
already emphasized, imply the second order field equations). With the change
of variable $X=\ln |{}\tan \frac{\alpha }{2}|{}$, $\alpha =2\arctan (e^{X})$
the BPS equations become the Cauchy-Riemann equations
\begin{eqnarray}
\partial _{x}X\pm \partial _{y}\Theta  &=&0,  \label{eq:CR_1} \\
\partial _{y}X\mp \partial _{x}\Theta  &=&0,  \label{eq:CR_2}
\end{eqnarray}%
which means that our fields are the real and imaginary parts of a locally
holomorphic function (complex field henceforth)
\begin{equation}
Z(w)=X(x,y)\pm \mathrm{i}\,\Theta (x,y),\quad w=x+\mathrm{i}\,y.
\end{equation}%
The main observables, the baryonic charge and the BPS charge, can be
expressed in terms of these new variables evaluated entirely on the boundary
$\partial R$ of the transverse rectangular domain. The baryonic charge reads
\begin{equation}
B=-\frac{p}{\pi }\int_{R}\sech^{3}X\sin \Theta \,dX\wedge d\Theta =-%
\frac{p}{2\pi }\oint_{\partial R}g[X]\sin \Theta \,d\Theta \ ,
\label{eq:baryon_charge_gX}
\end{equation}%
where the amplitude function $g[X]$ is defined as
\begin{equation}
g[X]\equiv 2\arctan (e^{X})+\sech X\tanh X\;.  \label{eq:def_gX}
\end{equation}%
Note that the quantity $B/p$ could be considered as the baryonic charge per
unit of length (this interpretation comes from the fact that neither the
baryonic charge density nor the energy density depend on $z$, one could also
say that the baryonic charge per unit of length in the $z$ direction is $p$%
). The BPS charge (namely, the topological charge which appears when
completing the squares in BPS style) is given by
\begin{equation}
Q_{BPS}=\int_{R}\text{sech}^{2}X\,dX\wedge d\Theta =\oint_{\partial R}\tanh
X\,d\Theta \ +2\pi n.
\end{equation}%
The fact that the physical solutions of our system are determined by a
locally holomorphic function $Z(w)$ (together with the main observables
being defined by boundary integrals) implies that the topological
information of the bulk is effectively encoded in the asymptotic behavior of
the fields on $\partial R$.

\subsection{On the boundary conditions and the non-vanishing baryonic charge}

For the $SU(2)$-valued field U to be single-valued under complete circuits
in the transverse domain R, the phase field $\Theta(x,y)$ must satisfy the
standard topological winding condition along the boundary $\partial R$:
\begin{equation}
\oint_{\partial R} d\Theta = 2\pi n \ , \quad n \in \mathbb{Z} \;.
\end{equation}
This condition naturally introduces the topological integer $n$. However,
evaluating the boundary integral for the baryonic charge,
\begin{equation}
B = -\frac{p}{2\pi} \oint_{\partial R} \Big( 2 \arctan(e^X) + \text{sech} X
\tanh X \Big) \sin \Theta \, d\Theta \;,
\end{equation}
one might naively worry that the integration of the $\sin\,\Theta$ term over
a full period could trivially vanish. This would indeed be the case if the
profile of the amplitude field X were constant along the boundary.

The underlying complex structure of the BPS equations prevents this
trivialization. Since the amplitude $X(x,y)$ and the phase $\Theta(x,y)$ are
the real and imaginary parts of the locally holomorphic function $Z(w)$,
they are harmonic conjugates deeply intertwined by the Cauchy-Riemann
equations. Any non-trivial winding of $\Theta$ along $\partial R$ strictly
requires a concurrent, non-constant spatial modulation of $X$ along the same
boundary.

By dynamically adapting $X$ to the phase variations, via imposing a
Riemann-Hilbert boundary condition as detailed in the next section and
Appendix \ref{app:solution}, it acts as a non-trivial weight $X\equiv
X(\Theta)$ along the contour. This functional dependence breaks the
odd-parity symmetry of the integrand, guaranteeing that the boundary
integral evaluates does not vanish.

\subsection{BPS baryonic-tube solutions via defect expansion}\label{BPSsol}
Since $Z(w)$ is a locally holomorphic function, multi-center
baryonic tube configurations can be directly constructed by inserting
logarithmic singularities (representing the topological defects) located at
arbitrary transverse positions, into a regular background field. Considering
the winding condition and the requirement of baryonic charge quantization,
we formulate the Riemann-Hilbert boundary problem on the unit disk (details
in Appendix \ref{app:solution}). By restricting the regular sector to a
physically motivated three-parameter polynomial truncation, the solution for
$n$ baryonic tubes takes the form
\begin{equation}
\mathcal{Z}(\zeta )=\mathcal{Z}_{top}(\zeta )+\mathcal{Z}_{reg}(\zeta
)=\sum_{j=1}^{n}\ln \left( \frac{\zeta -\zeta _{j}}{1-\bar{\zeta}_{j}\zeta }%
\right) +\frac{C_{0}+iB_{n}\zeta ^{n}-\overline{C_{0}}\zeta ^{2n}}{%
\prod_{j=1}^{n}(1-\bar{\zeta}_{j}\zeta )^{2}},
\end{equation}%
where $C_{0}=A_{0}+iB_{0}$, $A_{0},B_{0},B_{n}\in \mathbb{R}$ and $\zeta (w)$
is the complex coordinate on the unit disk. The first term on the right hand
side of the above equation contains relevant informations about the
topological defects while the second term is a regular holomorphic function
on the bulk that enforces the proper asymptotic behavior on the boundary.
From the homogeneous Riemann-Hilbert boundary condition (see Appendix \ref%
{app:solution})
\begin{equation}
a(\varphi )X_{\mathrm{reg}}(\varphi )-b(\varphi )\Theta _{\mathrm{reg}%
}(\varphi )=0,\quad a(\varphi )=\cos (\Theta _{top}(\varphi )),\quad
b(\varphi )=-\sin (\Theta _{top}(\varphi )),
\end{equation}%
we can explicitly see how the odd parity symmetry of the integrand is
broken, since we can substitute $X_{\mathrm{reg}}(\varphi )=\frac{b(\varphi )%
}{a(\varphi )}\Theta _{reg}(\varphi )$ in \eqref{eq:baryon_charge_gX},
yielding
\begin{align}
B& =-\frac{p}{2\pi }\oint_{\partial D}g[X]\sin \Theta d\Theta ,  \notag \\
& =-\frac{p}{2\pi }\int_{0}^{2\pi }g\left[ -\tan (\Theta _{top})\Theta _{reg}%
\right] \sin (\Theta _{top}+\Theta _{reg})\,[\Theta _{top}^{\prime }+\Theta
_{reg}^{\prime }]\,d\varphi ,  \notag
\end{align}%
preventing the baryonic charge from vanishing.

Regarding the BPS charge of the solutions, it can be computed rigorously
through Stokes' theorem by isolating the singularities. We excise small
disks $D_j$ of radius $\epsilon_j$ around the position of each tube $w_j$,
so the boundary integral over $R$ is compensated by the sum of the integrals
around the internal loops $C_j = \partial D_j$:
\begin{equation}
Q_{BPS} = \oint_{\partial R} \tanh X \, d\Theta - \sum_{j=1}^n
\lim_{\epsilon_j \to 0} \oint_{C_j} \tanh X \, d\Theta \ .
\end{equation}
Very close to the core of the $j$-th defect $w \to w_j$, setting $w - w_j =
\epsilon_j e^{\mathrm{i} \theta_j}$, the fields behave as
\begin{equation}
\Theta \approx \theta_j \ , \quad X \approx \ln \epsilon_j \ .
\end{equation}
In the limit $\epsilon_j \to 0$, we have $X \to -\infty$, which implies $%
\tanh X \to -1$. Therefore, the contribution from each defect loop is
precisely
\begin{equation}
\lim_{\epsilon_j \to 0} \oint_{C_j} \tanh X \, d\Theta = \int_{0}^{2\pi}
(-1) \, d\theta_j = -2\pi \ ,
\end{equation}
yielding the exact BPS charge formula
\begin{equation}
Q_{BPS} = \oint_{\partial R} \tanh X \, d\Theta + 2\pi n \;.
\end{equation}
In the case of the baryon charge $B$, the contribution from the internal
loops $C_j$ around the defects is strictly zero because $g[X]$ vanishes at
the cores,
\begin{equation}
\lim_{X \to -\infty} g[X] = \lim_{X \to -\infty} \Big( 2\arctan(e^X) + \sech X \tanh X \Big) = 0 \ .
\end{equation}

This confirms that the total baryonic charge arises exclusively from the
boundary integration over $\partial R$. Remarkably, these BPS solutions
permit the positions $w_{j}$ of the baryonic tubes to be completely
arbitrary (unlike what happens in the case of the baryonic tube which have
been already analyzed in the literature in \cite{OldSolutions}). This
reflects the physical feature derived in Section \ref%
{section_stress-energy_tensor}: the vanishing of the transverse pressures ($%
P_{x}=P_{y}=0$) guarantees that no static forces are exerted between the
topological defects, allowing equilibrium configurations for any spatial
arrangement of the baryonic tubes.

\subsection{Quantization condition and parameter space}

The phase kinematics guarantees that the boundary integral
scales directly with the topological defect index $n$, yielding a baryonic
charge of the form $B=-b\,n$, where $b$ is a continuous, real-valued
function of the background parameters $(A_{0},B_{0},B_{n})$. Thus, while the
topological phase winding $\oint_{\partial \mathbb{D}}d\Theta =2\pi n$
enforces the structural proportionality to $n$, the prefactor $b$ is
generally non-integer for generic parameter choices. The physical
requirement of exact integer quantization, $B=-np$ (with $p\in \mathbb{Z}$),
is realized by constraining the parameters to satisfy $b=p$. This matching
condition defines a closed two-dimensional extended manifold $\mathcal{M}%
\subset \mathbb{R}^{3}$, successfully promoting the topological winding into
a physically quantized baryonic number. On the other hand, the energy of the
baryonic tube which have been already analyzed in \cite{OldSolutions} grows
quadratically with the baryonic charge for large baryonic charges. Thus, as it will be discussed in the next section, at some point the novel solutions will be favoured over the old ones (for large enough baryonic charge).

\begin{figure}
\centering
\includegraphics[width=0.5\textwidth]{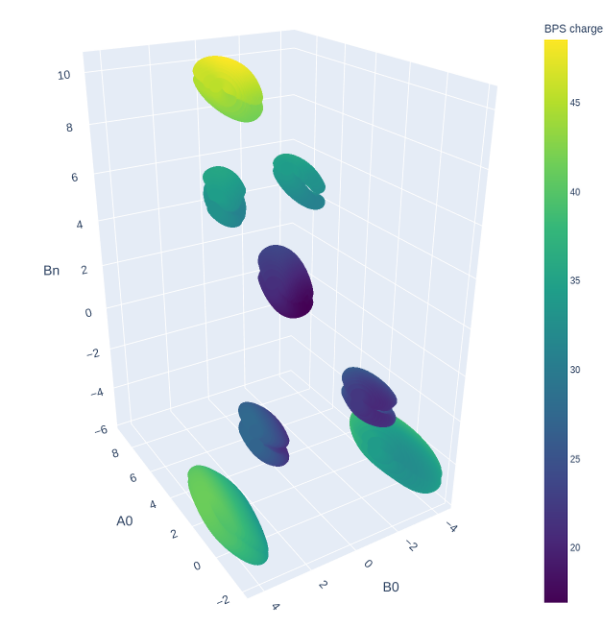}  %
\caption{The figure shows the numerically computed quantization
manifold (details in Appendix \ref{app:quantization_manifold_code}) to which the parameters $A_0$, $B_0$ and $B_n$ are constrained. The
coloring represents the value of the BPS charge. We can see that the minimum
values of the BPS charge are attained on the central “bubble”, and it
seems to increase as we move farther from it. The extended manifold also exhibits a reflection symmetry with respect to the $B_0=0$ plane, and changing the parameters of a configuration as $(A_0,B_0,B_n)\rightarrow (A_0,-B_0,B_n)$ yields no effect in the observables or the fields themselves.}  \label{fig:Qmanifold}
\end{figure}

This extended manifold $\mathcal{M}$ is found numerically and is visualized
as a collection of closed \textquotedblleft bubbles\textquotedblright , as
seen in Figure \ref{fig:Qmanifold}. Further numerical computations show
that this manifold exists for an arbitrary number of defects in arbitrary
positions. However, the closed surfaces are displaced and deformed in the
parameter space as the solitons move:
\begin{equation}
\mathcal{M}=\mathcal{M}[w_{1},w_{2},\dots ,w_{n}],
\end{equation}%
where $w_{j}=x_{j}+iy_{j}\in R$ represent the complex transverse positions
of the baryonic tubes. This implies that the background coefficients $%
(A_{0},B_{0},B_{n})$ possess an implicit dependence on the defect positions $%
w_{j}$. To systematically track a particular physical solutions given this
implicit dependence we utilize the exact BPS charge $Q_{\text{BPS}}$. Since
the topological index $n$ is invariant, $Q_{\text{BPS}}$ shifts smoothly
when the soliton positions change adiabatically. Thus, each point on the
two-manifold can be labelled with a $Q_{\text{BPS}}$, allowing us to trace
the parameters of specific physical configurations during spatial
deformations of the tube array.

\begin{figure}
    \centering
    \captionsetup{type=figure}

    % --- FILA 1: Fotogramas 1 y 2 ---
    \begin{subfigure}[c]{0.31\linewidth}
        \centering
        \includegraphics[width=\textwidth]{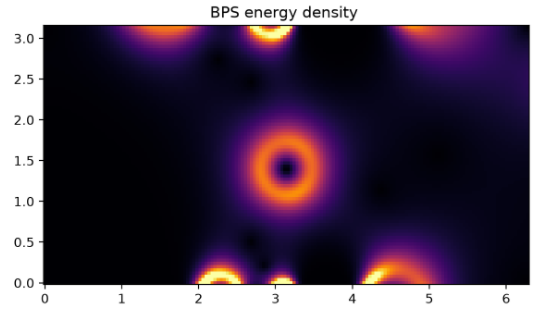}
    \end{subfigure}
    \begin{subfigure}[c]{0.31\linewidth}
        \centering
        \includegraphics[width=\textwidth]{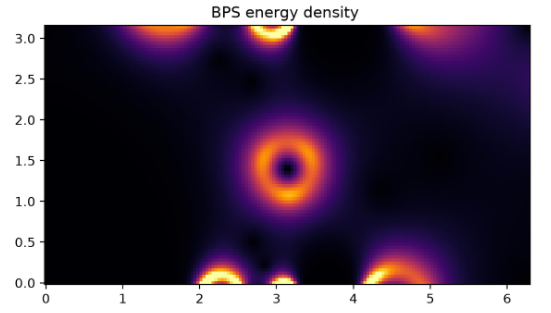}
    \end{subfigure}
    \begin{subfigure}[c]{0.31\linewidth}
        \centering
        \includegraphics[width=\textwidth]{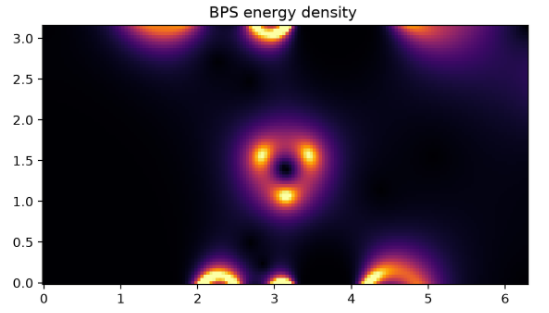}
    \end{subfigure}\hfill
    \begin{subfigure}[c]{0.04\linewidth}
        \centering
        \includegraphics[width=\textwidth]{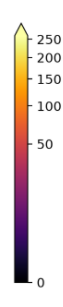}
    \end{subfigure}

    \vspace{0.8em}

    \begin{subfigure}[c]{0.31\linewidth}
        \centering
        \includegraphics[width=\textwidth]{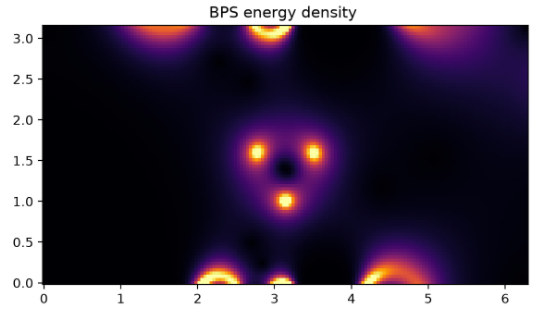}
    \end{subfigure}
    \begin{subfigure}[c]{0.31\linewidth}
        \centering
        \includegraphics[width=\textwidth]{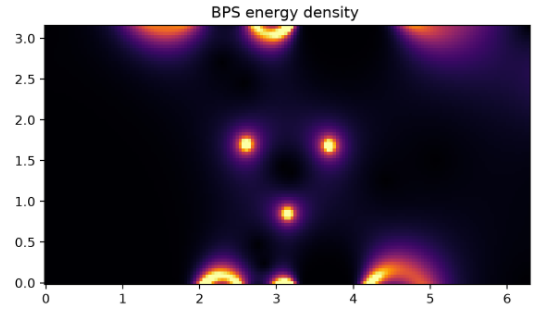}
    \end{subfigure}
    \begin{subfigure}[c]{0.31\linewidth}
        \centering
        \includegraphics[width=\textwidth]{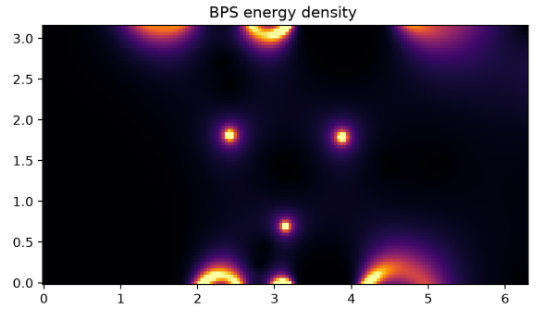}
    \end{subfigure}\hfill
    \begin{subfigure}[c]{0.04\linewidth}
        \centering
        \includegraphics[width=\textwidth]{bar1.png}
    \end{subfigure}

    \vspace{0.8em}

    \begin{subfigure}[c]{0.31\linewidth}
        \centering
        \includegraphics[width=\textwidth]{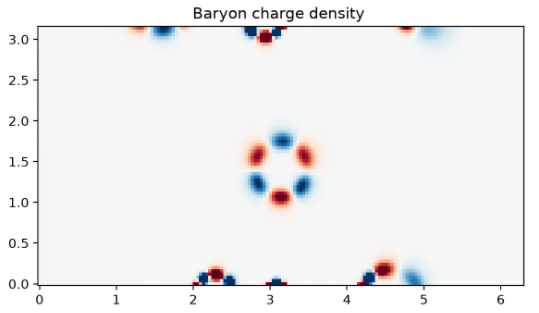}
    \end{subfigure}
    \begin{subfigure}[c]{0.31\linewidth}
        \centering
        \includegraphics[width=\textwidth]{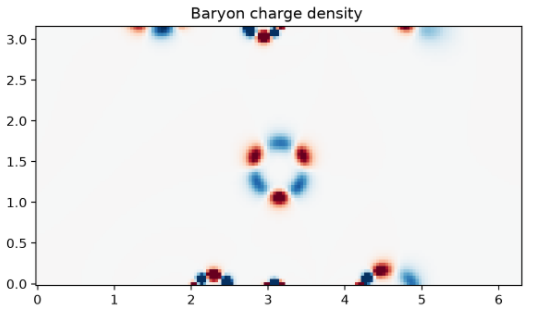}
    \end{subfigure}
    \begin{subfigure}[c]{0.31\linewidth}
        \centering
        \includegraphics[width=\textwidth]{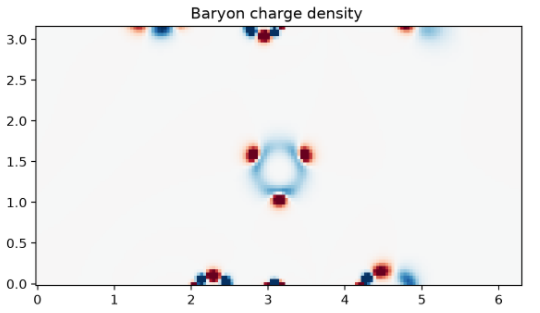}
    \end{subfigure}\hfill
    \begin{subfigure}[c]{0.04\linewidth}
        \centering
        \includegraphics[width=\textwidth]{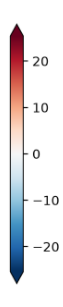}
    \end{subfigure}

    \vspace{0.8em}

    \begin{subfigure}[c]{0.31\linewidth}
        \centering
        \includegraphics[width=\textwidth]{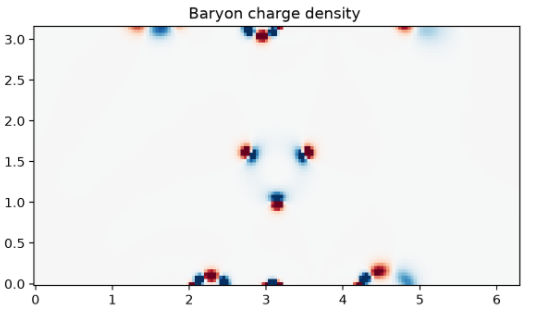}
    \end{subfigure}
    \begin{subfigure}[c]{0.31\linewidth}
        \centering
        \includegraphics[width=\textwidth]{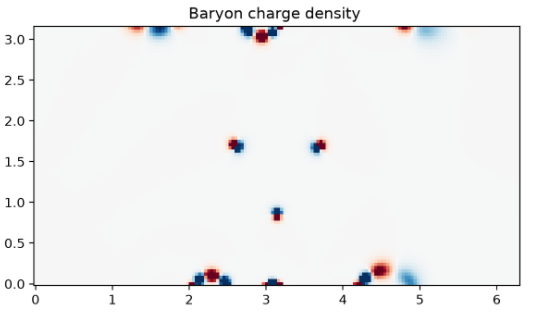}
    \end{subfigure}
    \begin{subfigure}[c]{0.31\linewidth}
        \centering
        \includegraphics[width=\textwidth]{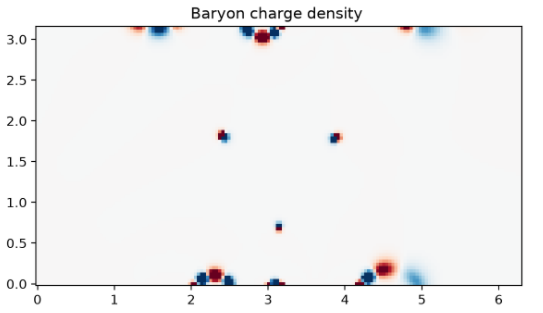}
    \end{subfigure}\hfill
    \begin{subfigure}[c]{0.04\linewidth}
        \centering
        \includegraphics[width=\textwidth]{bar2.png}
    \end{subfigure}
   \caption{These plots show the baryon charge density and BPS energy density (in units of $K/L^{2}$) for three solitons positioned at the vertices of an equilateral triangle, moving further apart. It is interesting to note that, when the solitons are close to each other, their charge and energy densities seem to form a ring around the ``centre of mass'', and as they move further apart, one can distinguish three peaks. On the other hand, while each tube core exhibits a localized, dipolar baryon density profile, the net integer baryon charge $B = -n p$ is integrated along the domain boundary through the regular component, whose modulation strictly reflects the total soliton number $n$ in the bulk.}
    \label{3solitons}
\end{figure}

\begin{figure}
    \centering
    \begin{subfigure}[b]{0.32\textwidth}
        \centering
        \includegraphics[width=\textwidth]{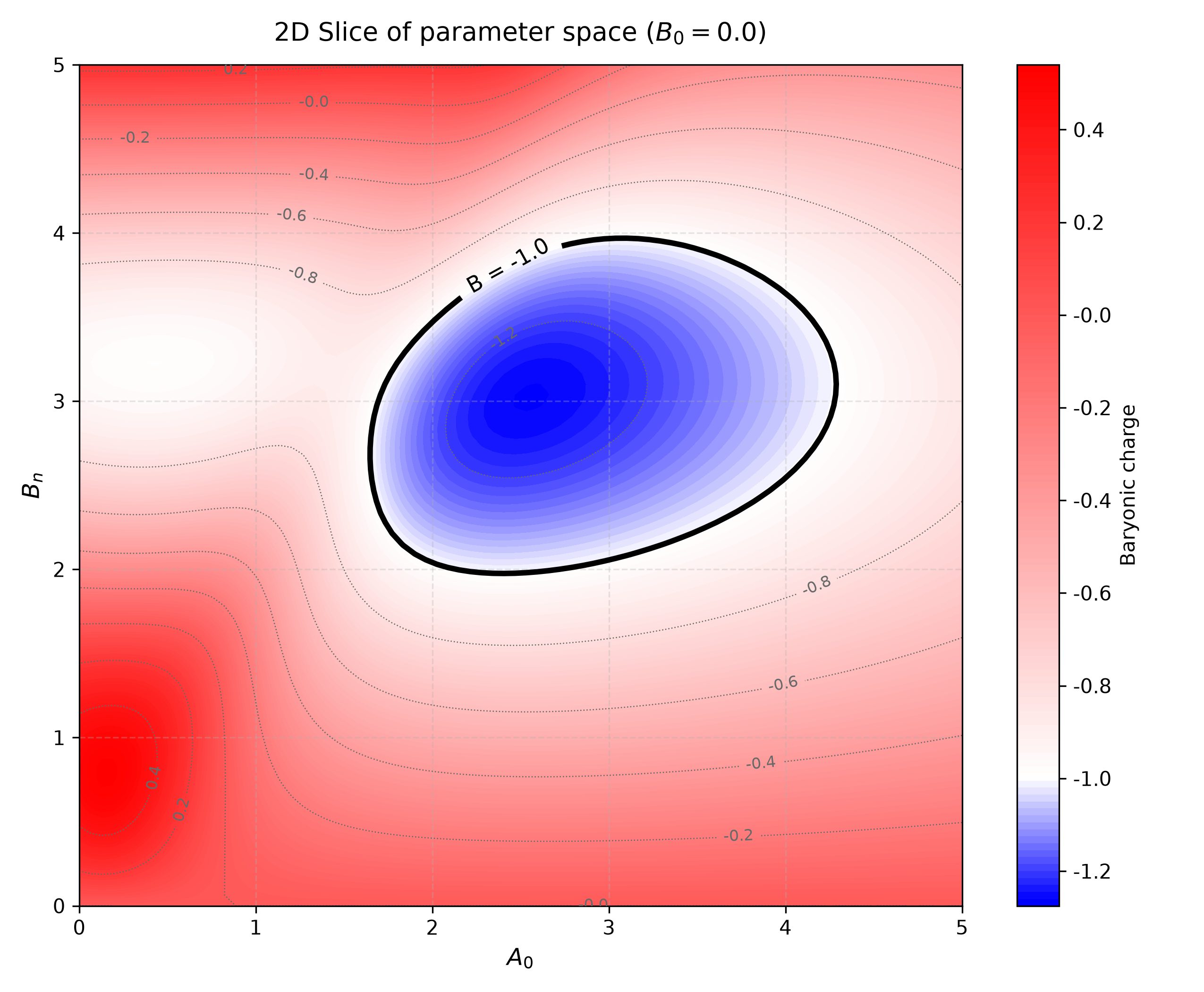}
    \end{subfigure}\hfill
    \begin{subfigure}[b]{0.32\textwidth}
        \centering
        \includegraphics[width=\textwidth]{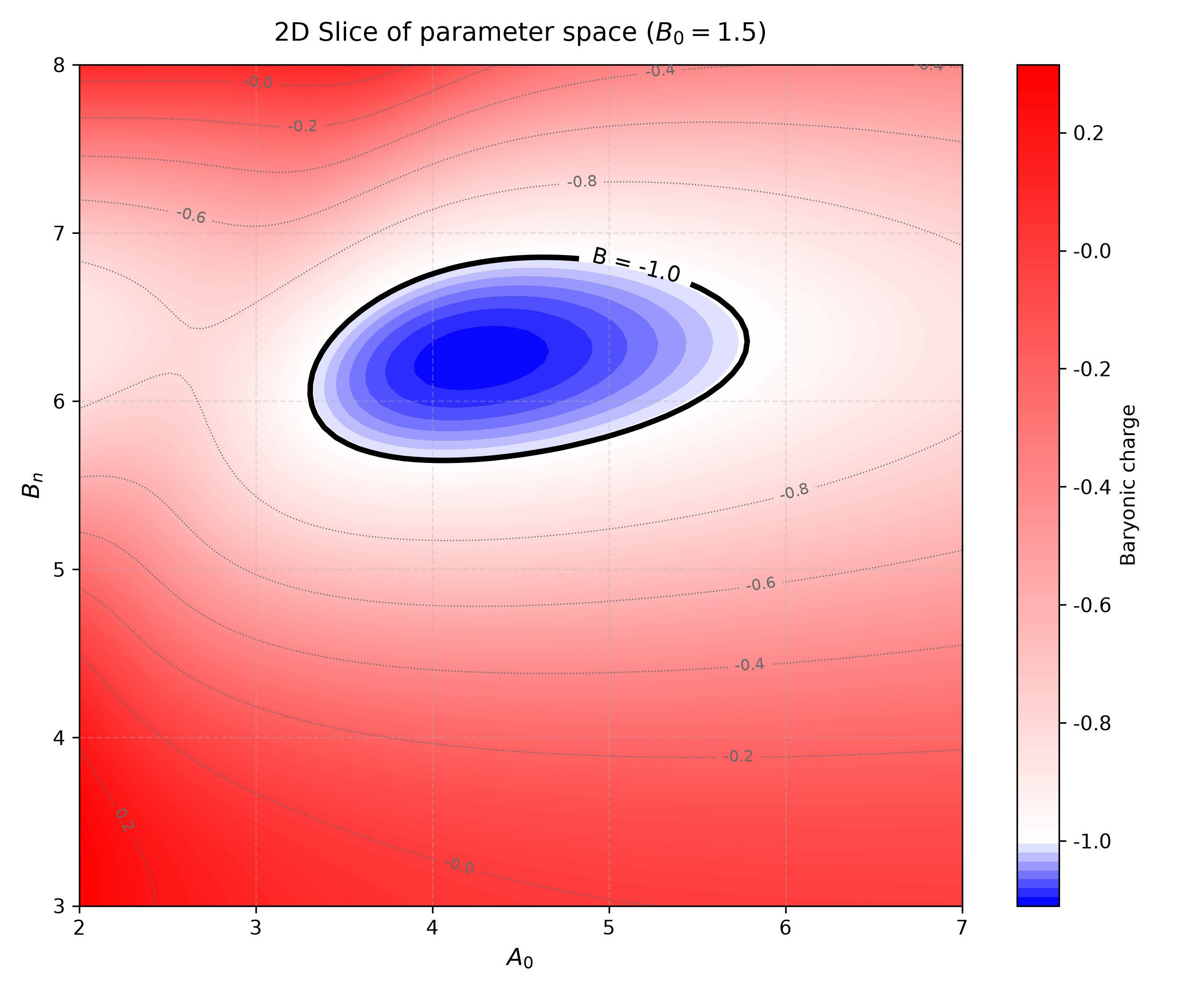}
    \end{subfigure}\hfill
    \begin{subfigure}[b]{0.32\textwidth}
        \centering
        \includegraphics[width=\textwidth]{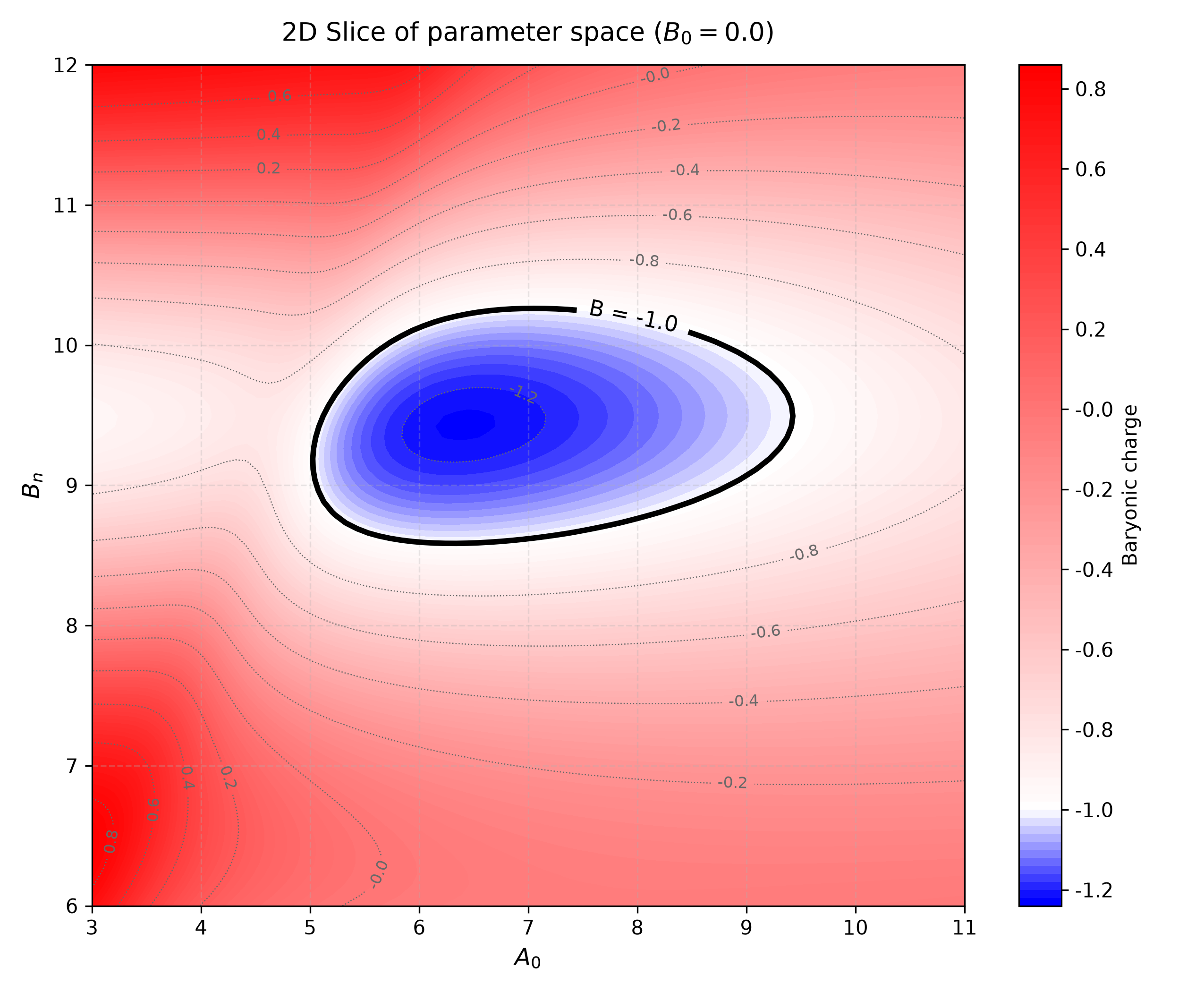}
    \end{subfigure}
    \caption{Slices in the $B_0$ direction of three bubbles from the quantization manifold in parameter space; the color map shows the values taken by the baryonic charge across these slices, representing the exact quantization with the color white. These plots correspond to (in reference to figure \ref{fig:Qmanifold}): the central bubble (left), the one right above the central bubble with positive $B_0$ (center) and the one at the top (right).}
    \label{2dslice}
\end{figure}

As seen in Figure \ref{2dslice}, our numerical calculations strongly suggest
that this quantization manifold serves as a frontier between two
qualitatively distinct regions of the parameter space, with the absolute
value of the baryonic charge exceeding the quantization value in the region
exterior to the bubbles and absent in the inner region.
A deeper understanding of this intricate inner structure is left to future
research. For the present purposes, by choosing points from the parameter
manifold $\mathcal{M}[w_1, \dots, w_n]$, Figure \ref{3solitons} illustrates
the resulting baryon charge density and BPS energy density for three
baryonic tubes as they are moved apart. The numerical plots clearly
demonstrate how the energy and charge density peaks track the arbitrary
positions $w_j$ of the defects, forming a ring-like state when close and
separating into distinct peaks at larger distances.

\section{Comparison with crystals of solitons}
It is highly instructive to compare these arbitrary-position BPS
configurations with the analytic, topologically non-trivial multi-soliton
crystals in the NLSM presented in \cite{OldSolutions}. In a similar setup,
starting from the rigidly structured ansatz
\begin{align}
\alpha = \alpha(x), \quad \Theta = qy, \quad \Phi =& p\left(\frac{t}{L_z} -
z\right), \quad q = 2v+1, \quad p, v \in \mathbb{N}, \quad p \neq 0, \\
A_\mu=&(u(x,y),0,0,-L_zu(x,y))\;,
\end{align}
the three NLSM equations reduce to a single ordinary differential equation
for $\alpha(x)$, which is equivalent to the first-order equation:
\begin{equation}
(\alpha^\prime)^2 = 2\left(E_0 - \frac{q^2}{4}\cos(2\alpha)\right) \equiv
(\eta(\alpha, E_0))^2, \quad E_0 > \frac{q^2}{4},
\end{equation}
while the Maxwell equations reduce to a single equation analogous to \ref%
{maxwell} with the potential $V=4L_z^2 K\sin^2(\alpha)\sin^2(qy)$. Imposing
the boundary conditions
\begin{equation}
\alpha(2\pi) - \alpha(0) = n\pi,\quad u(x,0)=u(x,\pi)=0,
\end{equation}
yields the baryon charge $B = -np$ and fixes $E_0$ in terms of $n$ via
elliptic integrals:
\begin{equation}
\pi\sqrt{2E_0 - q^2/2} = nK(m), \quad K(m) = \int_0^{\pi/2} \frac{d\alpha}{%
\sqrt{1 - m\sin^2\alpha}},
\end{equation}
where $m = -q^2 / (2E_0 - q^2/2)$ is the elliptic modulus. The energy per
unit of length along the $z$-axis for these crystals is:
\begin{equation}
\frac{E}{2\pi L_z} = 2Kn\sqrt{2E_0 - \frac{q^2}{2}} \bigg(2E(m) - K(m) %
\bigg) + \int_R (\boldsymbol{\nabla}u)^2 \,dxdy \;+ Kp^2\int_R
\sin^2(\alpha)\sin^2(qy)\left( \frac{2L_z u}{p}-1 \right)dxdy,
\end{equation}
with $E(m) = \int_0^{\pi/2} \sqrt{1 - m\sin^2\alpha} \, d\alpha$.

\begin{figure}
\centering
\includegraphics[width=0.5\linewidth]{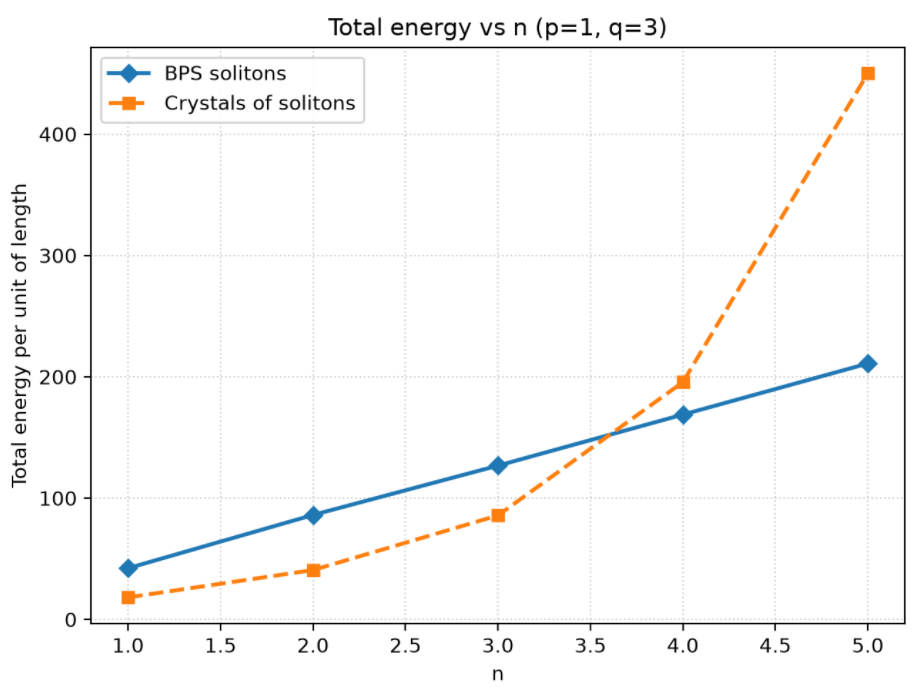}
\caption{figure}{This figure displays a comparison of the total energy per unit length (in units of $K$) between the analytic crystals of solitons and our BPS solitons.}
\label{fig:Comparison}
\end{figure}

While the soliton crystals described by these Jacobi elliptic functions
represent a rigid, equispaced periodic lattice, our novel BPS solutions
introduce a profound qualitative difference. Because our profiles are
governed by a locally holomorphic function $Z(w)$, the transverse pressures
strictly vanish ($P_{x}=P_{y}=0$), implying an absence of static
inter-soliton forces. This holomorphic mechanism is what allows the baryonic
tubes to be placed at completely arbitrary positions $w_{j}=x_{j}+iy_{j}$ in
the transverse plane $R$ without destabilizing the condensate. As
illustrated in the energy comparison figures (Figure \ref{fig:Comparison}),
the energy of our BPS solutions scales linearly with $n$ (maintaining a
constant base energy due to the non-topological $T_{0z}$ momentum term),
whereas the energy of the rigid soliton crystal scales quadratically with $n$%
. The intersection between both curves defines a critical topological
density (or crossover scale in baryon number) where an energetic crossing
between the equispaced crystal model and the BPS multi-tube family occurs.
This crossing indicates a potential structural transition in the preferred
ground-state geometry as the topological charge increases. Although the
non-topological momentum component prevents our BPS solutions from always
being the absolute energetic minimum, their capacity to sustain arbitrary
spatial arrangements $w_{j}$ provides a robust mechanism to study structural
transitions without the restrictive symmetries of standard crystalline
ansatze. Of course, one could wonder why the BPS solutions (being BPS) are not always favored over the old solutions (which are not BPS). The reason is that, as it has already been emphasized, the BPS bound leading to the present BPS equations is related to the action and not to the energy so
that there is a window in which the old solutions are favored over the novel
BPS ones.

\subsection{Electromagnetic field}
Introducing our BPS solution in \ref{maxwell} and solving numerically with
the boundary condition we can compute the electric and magnetic fields, as
well as the contributions from the Maxwell sector to the energy. The
non-vanishing components of the current are
\begin{align}
J_t=&\frac{2K}{L_{z}}\sin^2\alpha\sin^2\Theta (p-2L_z u), \\
J_z=& -2K \sin^2\alpha\sin^2\Theta (p-2L_z u),
\end{align}
while the components of the electric and magnetic fields are
\begin{align}
E_x&=-\partial_x u,\quad E_y=-\partial_y u,\quad E_z=0, \\
B_x&=\frac{1}{L^2L_z}\partial_y u,\quad B_y=-\frac{1}{L^2L_z}\partial_x
u,\quad B_z=0 .
\end{align}

\begin{figure}
    \centering

    % ==================== FILA SUPERIOR ====================
    % --- Grupo 1 (Superior) ---
    \begin{subfigure}[b]{0.49\linewidth}
        \centering
        \includegraphics[width=\linewidth]{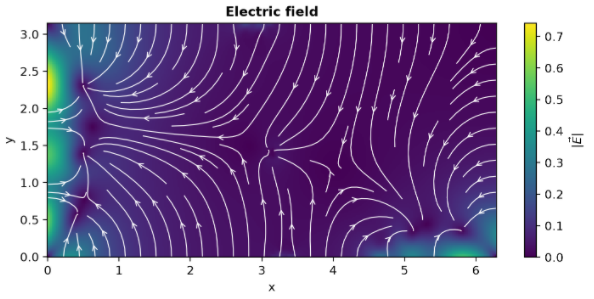}
    \end{subfigure}\hfill
    \begin{subfigure}[b]{0.49\linewidth}
        \centering
        \includegraphics[width=\linewidth]{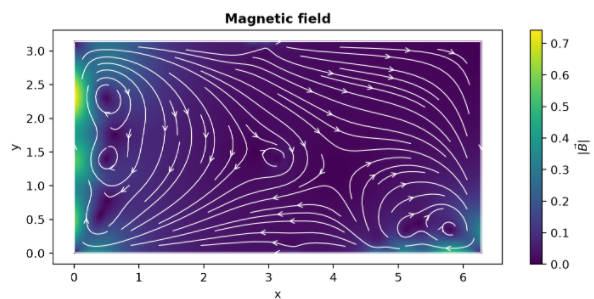}
    \end{subfigure}%

    \vspace{0.8em} % Separación vertical entre filas

    % ==================== FILA INFERIOR ====================
    % --- Grupo 1 (Inferior) ---
    \begin{subfigure}[b]{0.49\linewidth}
        \centering
        \includegraphics[width=\linewidth]{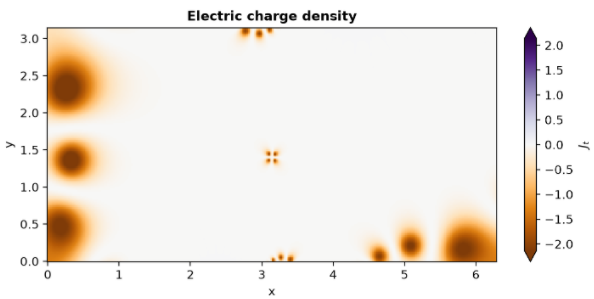}
    \end{subfigure}\hfill
    \begin{subfigure}[b]{0.49\linewidth}
        \centering
        \includegraphics[width=\linewidth]{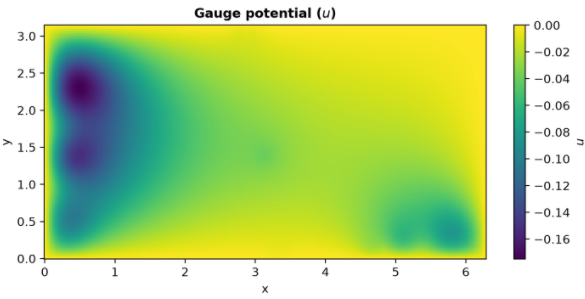}
    \end{subfigure}%
    % Separador entre Grupo 1 y Grupo 2
    \vspace{0.8em}
    % --- Grupo 2 (Superior) ---
    \begin{subfigure}[b]{0.49\linewidth}
        \centering
        \includegraphics[width=\linewidth]{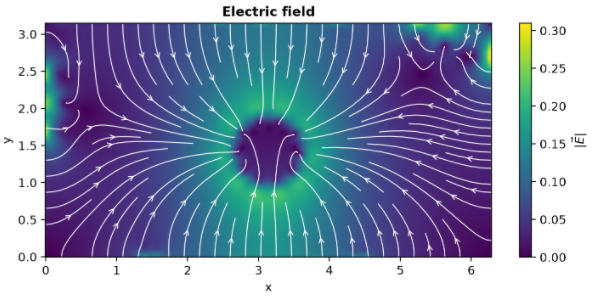}
    \end{subfigure}\hfill
    \begin{subfigure}[b]{0.49\linewidth}
        \centering
        \includegraphics[width=\linewidth]{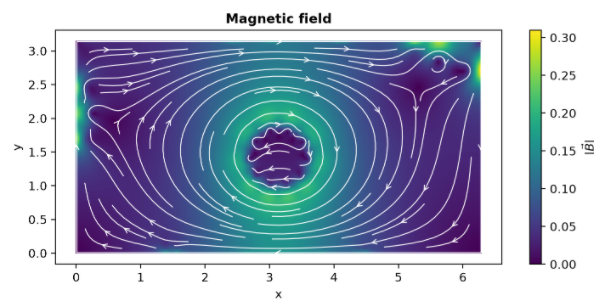}
    \end{subfigure}

    \vspace{0.8em}
    % --- Grupo 2 (Inferior) ---
    \begin{subfigure}[b]{0.49\linewidth}
        \centering
        \includegraphics[width=\linewidth]{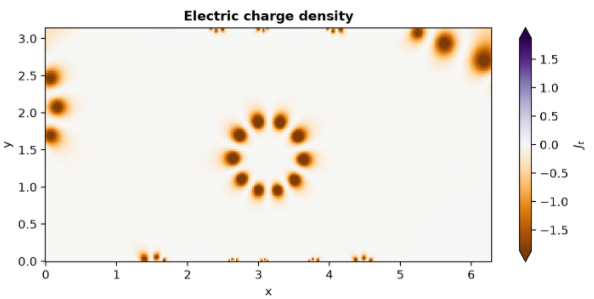}
    \end{subfigure}\hfill
    \begin{subfigure}[b]{0.49\linewidth}
        \centering
        \includegraphics[width=\linewidth]{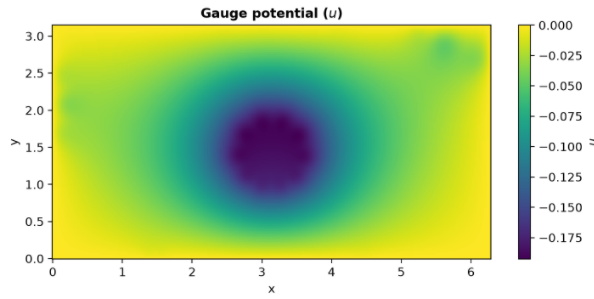}
    \end{subfigure}

    \vspace{1em}
    \caption{This figure displays the electric and magnetic fields (in units of $1/L^2$), as well as the electric charge density (in units of $K/L$) and the Maxwell potential $u(x,y)$ (in units of $1/L$), for configurations with $n=2$ (top) and $n=5$ (bottom).}
    \label{fig:campos_comparacion}
\end{figure}

To further our physical interpretation of our BPS solitons, we provide in
Figure \ref{fig:campos_comparacion} the behavior of the electromagnetic
field corresponding to two different regimes: one in which the BPS solitons
are energetically favourable in comparison with the crystals of solitons,
and another in which they are not, as seen in Figure \ref{fig:Comparison}.

We can see that, in the first regime with low $n$ (where the old solutions
are energetically favored), the electric charge and therefore the electric
and magnetic fields of the novel solutions are spread out throughout the
boundary of the region, while for higher $n$ when the BPS solitons are the
most favourable, the electric charge and the fields seem to concentrate on
the position of the solitons.

\section{Conclusions and perspectives}
In the present paper, we have constructed for the first time a BPS system in
$(3+1)$-dimensional chiral perturbation theory that yields second-order
field equations in the case of baryonic tubes. The baryonic tubes saturating
this bound describe configurations whose energy-density peaks can be placed
at arbitrary positions in the transverse plane due to the vanishing of the
transverse stress tensor (the inclusion of the electromagnetic field does
not spoil these BPS features). While the primary focus of this work has been
on the leading non-linear sigma model (NLSM) action, the generalization to
include potential terms and their immediate physical consequences are
presented in Appendix \ref{app:potential_term}. The framework
introduced here can have a significant impact on the analysis of CPT,
extending previous analytical techniques developed for scalar field theories
and non-linear sigma models~\cite{Alonso-Izquierdo2023,Alonso-Izquierdo2026}
and, more generically, on QCD under extreme conditions of non-vanishing
baryon density, strong magnetic fields, and low temperatures regime where
lattice QCD faces severe sign-problem limitations and standard analytical
techniques are ineffective. For instance, our approach enables a precise
determination of the magnetic flux required to increase the baryonic charge
of the condensate, a task that remains challenging with other methods.
Another promising application is the study of quark couplings to these
gauged, magnetized BPS baryonic configurations via Dirac-like equations in $%
SU(2)$ chiral background fields \cite%
{quarkskyrme1,quarkskyrme2,quarkskyrme3,quarkskyrme31,quarkskyrme4}. A
further very interesting question is the role of the subleading terms in
Chiral Perturbation Theory in the analysis of these novel BPS tubes. We plan
to report on these relevant issues in future publications.

\subsection*{Acknowledgements}
This work has been funded by Fondecyt grants 1240048 and by Grant ANID EXPLORACION 13250014. The Centro de Estudios Cientificos (CECs) is funded by the Chilean Government through the Centers of Excellence Base Financing Program of Conicyt. P.~P. gladly acknowledge support from Charles University Research Centres (UNCE 24/SCI/016).

\begin{appendices}

\section{On the construction of the solution}\label{app:solution}
In this appendix we provide a detailed explanation of the analytical methods utilized in constructing a solution of the form $\mathcal{Z}(\zeta)=\mathcal{Z}_{top}(\zeta)+\mathcal{Z}_{reg}(\zeta)$, where the first component introduces the topological defects, and the second component is a regular function that enforces the proper behavior on the boundary.

We begin by performing a conformal mapping of our physical space from a rectangle to the unit disk and proposing a topological component of the complex field whose phase winds around the boundary by $2n$, with $n$ being the number of defects. Then we formulate the Riemann-Hilbert boundary problem on the unit disk, following the classic formulation of N.~I. Muskhelishvili~\cite{muskhelishvili1958}, to construct the regular component and fulfill the non-vanishing and quantization of the baryonic charge.
\subsection{Conformal mapping to the unit disk}
By composing several transformations on the cartesian coordinate $w=x+iy$, we obtain the conformal transformation
\begin{equation}
    \zeta(w)=\frac{\sn\left((\frac{w}{\pi}-1)\mathbf{K},k\right)-i}{\sn\left((\frac{w}{\pi}-1)\mathbf{K},k\right)+i},
\end{equation}
where $\mathbf{K}$ denotes a complete elliptic integral of the first kind and $k=1/\sqrt{2}$ is the elliptic module. This transformation maps the rectangle $0\leq x\leq 2\pi$, $0\leq y \leq\pi$ to the unit disk
\begin{equation}
    0\leq r \leq 1,\quad 0\leq \varphi \leq 2\pi,
\end{equation}
where $r=|\zeta|$ and $\varphi=\arg(\zeta)$.The Jacobian of the transformation is
\begin{equation}
    \left| \frac{dw}{d\zeta}\right|^2=\frac{2\pi^2}{\mathbf{K}^2}\frac{1}{|1+\zeta^2|\cdot|(1-\zeta)^2+k^2(1+\zeta)^2|}
\end{equation}
and the inverse is
\begin{equation}
    \left| \frac{d\zeta}{dw} \right|^2 = \frac{4\mathbf{K}^2}{\pi^2} \frac{|\operatorname{cn}(s, k)|^2 |\operatorname{dn}(s, k)|^2}{|\operatorname{sn}(s, k) + i|^4}, \quad  s = \left(\frac{w}{\pi} - 1\right)\mathbf{K}
\end{equation}
Working on the unit disk has several advantages: \textit{Firstly}, it smoothes the boundary, encoding the corner singularities entirely within the Jacobian of the map. \textit{Secondly}, mapping to the unit disk aligns the boundary with the phase winding, making the complex logarithm a natural ansatz. \textit{Thirdly}, as we shall see later, boundary value problems such as the Riemann-Hilbert problem have a well established formulation on the unit disk.

\subsection{Logarithmic singularities}
As remarked above, the boundary winding condition and circular symmetry naturally lead to a logarithmic ansatz. The simplest choice is $n\ln \zeta$, which accumulates a net phase of $2\pi n$ along the boundary by placing a topological defect at the origin. This ansatz is generalized straightforwardly to defects at arbitrary positions $\zeta_j$ via a finite Blaschke product:
\begin{equation}
    \mathcal{B}(\zeta)=\prod_{j=1}^n \frac{\zeta - \zeta_j}{1 - \bar{\zeta}_j \zeta}\;,
\end{equation}
which introduces a zero at the position of each defect while maintaining a constant unit module at the boundary. This way, the topological component of the general solution is:
\begin{equation}
    \mathcal{Z}_{top}(\zeta)=\ln\big( \mathcal{B}(\zeta)\big)=X_{top}+i\Theta_{top}.
\end{equation}
While this component effectively places topological defects at arbitrary positions accumulating a net phase of $2\pi n$ and having BPS charge quantized in terms of $n$, it does not manage to translate this topological number into the baryonic charge, which vanishes due to the lack of modulation of the amplitude field $X$. This is corrected by imposing a Riemann-Hilbert boundary condition that produces a dynamical coupling between the amplitude and the phase of the solution via the construction of a regular background component, and then correctly tunning the parameters of this regular component.

\subsection{The Riemann-Hilbert problem on the unit disk}
Let $S^+ = \{ |\zeta| < 1 \}$, $S^- = \{ |\zeta| > 1 \}$, and $L = \{ t = e^{i\varphi} : |t| = 1 \}$. The classical Riemann-Hilbert problem seeks the regular background function $\mathcal{Z}_{\text{reg}}(\zeta) = X_{\text{reg}}(\zeta) + i\Theta_{\text{reg}}(\zeta)$, strictly holomorphic in $S^+$ and continuous on $S^+ \cup L$, satisfying on $L = \{t = e^{i\varphi} : |t|=1\}$ the boundary condition:
\begin{equation}\label{eq:app_rh_boundary}
a(\varphi) X_{\text{reg}}(t) - b(\varphi) \Theta_{\text{reg}}(t) = 0,
\end{equation}
where the continuous coefficients $a(\phi) = \cos(\Theta_{\text{top}}(\phi))$ and $b(\phi) = -\sin(\Theta_{\text{top}}(\phi))$ are determined by the boundary values of the topological phase, satisfying the Hölder condition, with $a^2 + b^2 \neq 0$. We then extend $\mathcal{Z}_{\text{reg}}$ to the whole plane via Schwarz reflection:
\begin{equation}
    Z(\zeta)=
    \begin{cases}
        \mathcal{Z}_{\text{reg}}(\zeta)& \text{for } \zeta \in S^+,\\
        \\
        \overline{\mathcal{Z}_{\text{reg}}\left(\frac{1}{\bar{\zeta}}\right)}& \text{for } \zeta \in S^- .
    \end{cases}
\end{equation}
Since $t\bar{t} = 1$ on $L$, the boundary limits satisfy $Z^+(t) = \mathcal{Z}_{\text{reg}}^+(t)$ and $Z^-(t) = \overline{\mathcal{Z}_{\text{reg}}^+(t)}$, which express the real and imaginary parts as $X(t) = \frac{1}{2}(Z^+ + Z^-)$ and $\Theta(t) = \frac{1}{2i}(Z^+ - Z^-)$. Substituting these into Eq.~\eqref{eq:app_rh_boundary} reduces the condition to the standard Riemann jump problem:
\begin{equation}
    Z^+(t) = G(t)Z^-(t) + g(t), \quad G(t) = -\frac{a(t) - ib(t)}{a(t) + ib(t)}, \quad g(t) = \frac{2c(t)}{a(t) + ib(t)}.
    \label{eq:app_riemann_jump}
\end{equation}
The solvability is governed by the topological index $\kappa = \text{ind } G(t) = \frac{1}{\pi}[\arg(a - ib)]_L$, which is an even integer for continuous real coefficients. For the homogeneous case ($c=0$), the fundamental non-vanishing solution $\Omega(\zeta)$ satisfying $\Omega^+(t) = G(t)\Omega^-(t)$ is constructed as
\begin{equation}
    \Omega(\zeta) =
    \begin{cases}
        e^{\Gamma(\zeta)}, & |\zeta| < 1, \\
        \zeta^{-\kappa} e^{\Gamma(\zeta)}, & |\zeta| > 1,
    \end{cases}
    \quad \text{with} \quad
    \Gamma(\zeta) = \frac{1}{2\pi i} \oint_L \frac{\ln [t^{-\kappa} G(t)]}{t - \zeta} dt.
\end{equation}
For $\kappa \geq 0$, the general homogeneous solution is $Z(\zeta) = \Omega(\zeta) P_\kappa(\zeta)$, where $P_\kappa(\zeta) = \sum_{k=0}^{\kappa} C_k \zeta^k$ ($C_k \in \mathbb{C}$) is an arbitrary polynomial of degree $\kappa$ with complex coefficients. Enforcing the reality and reflection symmetry $Z(\zeta) = \overline{Z(1/\bar{\zeta})}$ on $S^+$, alongside the identity $\overline{\Omega(1/\bar{\zeta})} = \zeta^{-\kappa} \Omega(\zeta)$ (see \cite{muskhelishvili1958}), imposes the algebraic constraint
\begin{equation}
    C_k = -\bar{C}_{\kappa - k}, \quad \forall k \in \{0, 1, \dots, \kappa\},
\end{equation}
restricting the solution to exactly $\kappa + 1$ independent real parameters.
\subsection{Construction of the full BPS solution}
We now apply the formulation of the (homogeneous) Riemann-Hilbert problem presented to construct the regular part of the solution. On the boundary $L$, the real fields $X = \mathrm{Re}[\mathcal{Z}]$ and $\Theta = \mathrm{Im}[\mathcal{Z}]$ satisfy the linear Riemann-Hilbert boundary condition
\begin{align}
    a(\varphi)\left(\cancelto{0}{X_{\mathrm{top}}(\varphi)} + X_{\mathrm{reg}}(\varphi)\right) - b(\varphi)\left(\Theta_{\mathrm{top}}(\varphi) + \Theta_{\mathrm{reg}}(\varphi)\right) &= c(\varphi), \\
    a(\varphi)X_{\mathrm{reg}}(\varphi) - b(\varphi)\Theta_{\mathrm{reg}}(\varphi) &= c(\varphi) + b(\varphi)\Theta_{\mathrm{top}}(\varphi)
\end{align}
Tuning the boundary function to $c(\varphi) =- b(\varphi)\Theta_{\mathrm{top}}(\varphi)$, yields the classical homogeneous Riemann-Hilbert boundary value problem for the regular field component
\begin{equation}
    a(\varphi)X_{\mathrm{reg}}(\varphi) - b(\varphi)\Theta_{\mathrm{reg}}(\varphi) = 0.
\end{equation}
To match the symmetry of the bulk defects on the boundary, we select real continuous boundary as
\begin{equation}
    a(\varphi) = \cos\left(\Theta_{\mathrm{top}}(\varphi)\right), \quad b(\varphi) = -\sin\left(\Theta_{\mathrm{top}}(\varphi)\right).
\end{equation}
With this choice, the complex boundary jump function $G(t)$ is
\begin{equation}
    G(t) = -\frac{a(t) - ib(t)}{a(t) + ib(t)} = -\frac{\cos\Theta_{\mathrm{top}} + i\sin\Theta_{\mathrm{top}}}{\cos\Theta_{\mathrm{top}} - i\sin\Theta_{\mathrm{top}}} = -e^{2i\Theta_{\mathrm{top}}(\varphi)} = -\mathcal{B}(t)^2,
\end{equation}
and the topological index turns out to be $\kappa=2n$. Then the fundamental solution $\Omega(\zeta) = e^{\Gamma(\zeta)}$ of the homogeneous boundary jump condition $Z^+(t) = G(t)Z^-(t)$ is determined by
\begin{equation}
    \Gamma(\zeta) = \frac{1}{2\pi i} \oint_L \frac{\ln\left[t^{-2n} G(t)\right]}{t - \zeta} dt = \frac{1}{2\pi i} \oint_L \frac{\ln\left[-t^{-2n} \mathcal{B}(t)^2\right]}{t - \zeta} dt.
\end{equation}
Expanding the logarithm and inserting the explicit formula of the Blaschke product yields
\begin{equation}
    \ln\left[-t^{-2n} \mathcal{B}(t)^2\right] = i\pi + 2\sum_{j=1}^n \ln\left(1 - \frac{\zeta_j}{t}\right) - 2\sum_{j=1}^n \ln\left(1 - \bar{\zeta}_j t\right).
\end{equation}
We identify the sectionally holomorphic components
\begin{align}
    V^-(\zeta) &= i\pi + 2\sum_{j=1}^n \ln\left(1 - \frac{\zeta_j}{\zeta}\right), \quad  S^- \ (| \zeta | > 1), \\
    V^+(\zeta) &= -2\sum_{j=1}^n \ln\left(1 - \bar{\zeta}_j \zeta\right), \quad  S^+ \ (| \zeta | < 1).
\end{align}
By applying the Plemelj-Sokhotski projections, the Cauchy integral selects the interior holomorphic component $\Gamma(\zeta) = V^+(\zeta) + \ln C$. Exponentiating this result, we find the exact fundamental function
\begin{equation}
    \Omega(\zeta) = C \prod_{j=1}^n \left(1 - \bar{\zeta}_j \zeta\right)^{-2},
\end{equation}
and the general solution is given by the product of the fundamental function $\Omega(\zeta)$ with an arbitrary polynomial $P_{2n}(\zeta)$ of degree $2n = \kappa$
\begin{equation}
    \mathcal{Z}_{\text{reg}}=\Omega(\zeta) P_{2n}(\zeta) = \frac{C_0 + C_1\zeta + \dots + C_{2n}\zeta^{2n}}{\prod_{j=1}^n (1 - \bar{\zeta}_j \zeta)^2}, \quad C_k = -\bar{C}_{2n-k}.
\end{equation}
 In general, this space of solutions possesses $2n + 1$ real independent parameters. To construct a transparent, explicit physical family of solutions valid for any number of defects $n \in \mathbb{Z}_{>0}$, we consider a symmetric three-parameter truncation of $P_{2n}(\zeta)$ by retaining only the terms of degree $0, n,$ and $2n$: $$P_{2n}(\zeta) \to C_0 + i B_n \zeta^n - \bar{C}_0 \zeta^{2n}, \quad C_0 = A_0 + i B_0,$$ where $A_0, B_0, B_n \in \mathbb{R}$. The resulting three-parameter family of BPS solutions reads:
\begin{equation}
    \mathcal{Z}(\zeta)=\sum_{j=1}^{n} \ln \left( \frac{\zeta - \zeta_j}{1 - \bar{\zeta}_j \zeta} \right) + \frac{C_0 +iB_n\zeta^n - \overline{C_{0}}\zeta^{2n}}{\prod_{j=1}^n (1 - \bar{\zeta}_j \zeta)^2}.
\end{equation}

\section{Generalization to include potential terms}
\label{app:potential_term}

In this appendix, we generalize the BPS equations discussed in the main text by incorporating specific potential terms in the chiral Lagrangian (see, e.g., [40–42] for analogous structures in similar topological contexts). To this end, we consider the following modified first-order BPS system:
\begin{align}
\partial_x \alpha \pm \sin \alpha \, \partial_y \Theta - W_1 &= 0 \;, \label{lambda0_extended_BPSeq1} \\
\partial_y \alpha \mp \sin \alpha \, \partial_x \Theta - W_2 &= 0 \;, \label{lambda0_extended_BPSeq2}
\end{align}
where $W_{1}=W_{1}(\alpha,\Theta)$ and $W_{2}=W_{2}(\alpha,\Theta)$ are potential-dependent functions. By applying cross-differentiation to match the second-order field equations of the non-linear sigma model (ensuring no unwanted first-derivative terms survive), we take the partial derivative of Eq.~\eqref{lambda0_extended_BPSeq1} with respect to $x$ and add it to the derivative of Eq.~\eqref{lambda0_extended_BPSeq1}  with respect to $y$, obtaining:
\begin{align}
\Delta\alpha-\sin\alpha\cos\alpha(\nabla\Theta)^{2}-\frac{1}{2}\frac{\partial
}{\partial\alpha}\Big(W_{1}^{2}+W_{2}^{2}\Big) &  \nonumber\label{alpha2}\\
-\left[  \cos\alpha W_{2}+\sin\alpha\frac{\partial W_{2}}{\partial\alpha
}+\frac{\partial W_{1}}{\partial\Theta}\right]  \frac{\partial\Theta}{\partial
x}+\left[  \cos\alpha W_{1}+\sin\alpha\frac{\partial W_{1}}{\partial\alpha
}-\frac{\partial W_{2}}{\partial\Theta}\right]  \frac{\partial\Theta}{\partial
y}= &  0\;,
\end{align}

Similarly, taking cross-derivatives combined with $\sin \alpha$ yields the second-field equation condition for $\Theta$. For the field equations to remain free of anomalous first-derivative couplings, the functions $W_{1}$ and $W_{2}$ must satisfy the pair of differential constraints:
\begin{align}
\frac{\partial (\sin \alpha \, W_1)}{\partial \alpha} - \frac{\partial W_2}{\partial \Theta} &= 0 \;, \\
\frac{\partial (\sin \alpha \, W_2)}{\partial \alpha} + \frac{\partial W_1}{\partial \Theta} &= 0 \; .
\end{align}

Rescaling the functions as $\widetilde{W}_{1}=\sin\alpha\,W_{1}$ and $\widetilde{W}_{2}=\sin\alpha\,W_{2}$, and implementing the logarithmic field transformation $X(\alpha)=\ln \tan(\alpha/2)$ (such that $\sin\alpha\,\partial/\partial\alpha=\partial/\partial X$), the system maps directly onto the Cauchy-Riemann equations:
\begin{align}
\frac{\partial \tilde{W}_1}{\partial X} - \frac{\partial \tilde{W}_2}{\partial \Theta} &= 0 \;, \\
\frac{\partial \tilde{W}_2}{\partial X} + \frac{\partial \tilde{W}_1}{\partial \Theta} &= 0 \;.
\end{align}

This proves that $\widetilde{W}_{1}$ and $\widetilde{W}_{2}$ constitute the real and imaginary parts of an analytic function $V(Z)$ of the complex variable $Z=X+\ii \Theta$. Requiring that the potential invariant $\widetilde{W}_{1}^{2} + \widetilde{W}_{2}^{2}=f(\alpha)^{2}$ depends solely on the amplitude $\alpha$ leads to the most general separable solution:
\begin{equation}\label{f_extra_term}
f(\alpha)^2 = A^2 \frac{(1 - \cos \alpha)^{a-1}}{(1 + \cos \alpha)^{a+1}} = A^2 \frac{(\tan(\alpha/2))^{2a}}{\sin^2 \alpha}\;,
\end{equation}
where A and a are real constants. The resulting effective potential term in the modified $\alpha$-equation of motion evaluates to:
\begin{equation}\label{effective_potential}
G(A, a, \alpha) \equiv \frac{1}{2} \frac{\partial}{\partial \alpha} (W_1^2 + W_2^2) = C \frac{(\tan(\alpha/2))^{2a-1}}{\sin^3 \alpha} (a - \cos \alpha) \;.
\end{equation}

\begin{figure}
\centering
\includegraphics[width=0.80\textwidth,angle=0]{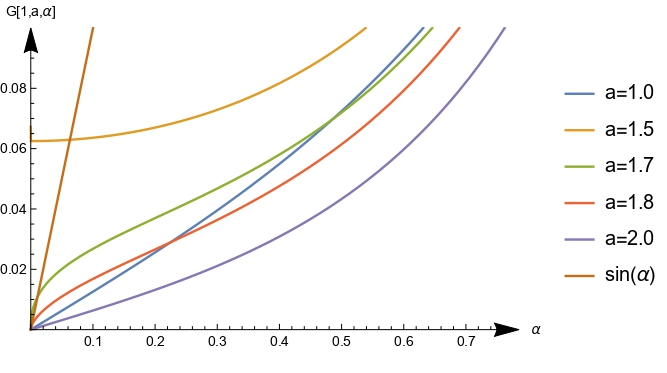}
\caption{The $G(1,a,\alpha)$ is a function of $\alpha$ for different values of $a$. It can be seen that neither of them takes the form of a ``mass term'' with $\sin \alpha$.}%
\label{kind_of_mass_fig}%
\end{figure}

To analyze whether these expressions can simulate a standard mass term (proportional to $\sin\alpha$), we evaluate the function $G(1,a,\alpha)$ numerically. As illustrated in Fig.~\ref{kind_of_mass_fig}, which plots $G(1,a,\alpha)$ as a function of $\alpha$ for various values of the parameter $a$, none of the curves strictly replicate the simple linear behavior of a standard mass term $\sin\alpha$ across the entire domain, highlighting the rich non-linear structure of these potential corrections. In the weak-field limit ($\alpha \ll 1$), choosing $a=2$ approximates a mass-like behavior $V'(\alpha)\approx\frac{C}{16}\sin\alpha$ , allowing the identification $C=64m^{2}/K$.
To construct the modified BPS energy bound, we rewrite the free energy functional by completing squares:
\begin{equation}
E = \int d^2x \left\{ \frac{1}{2}(\partial_x \alpha \pm \sin \alpha \partial_y \Theta - W_1)^2 + \frac{1}{2}(\partial_y \alpha \mp \sin \alpha \partial_x \Theta - W_2)^2 \pm \sin \alpha (\partial_y \alpha \partial_x \Theta - \partial_x \alpha \partial_y \Theta) \pm \vec{\nabla}\cdot\vec{J} \right\}\;,
\end{equation}
where $\vec{J}=(J_{x},J_{y})$ is defined via the current integration conditions $\partial_{\alpha}J_{x} = \pm W_{1}$, $\partial_{\Theta}J_{y} =- \sin\alpha W_{2}$. Solving these relations yields the explicit current components:
\begin{equation}
J_x = \pm \frac{(\tan(\alpha/2))^a \cos(a\Theta)}{a} \;, \quad  \quad J_y = \frac{(\tan(\alpha/2))^a \sin(a\Theta)}{a} \;.
\end{equation}

\begin{figure}
    \centering
    \captionsetup{type=figure}
     % FILA 1: Imágenes A y B
     \begin{subfigure}[b]{0.45\textwidth}
         \centering
         \includegraphics[width=\textwidth]{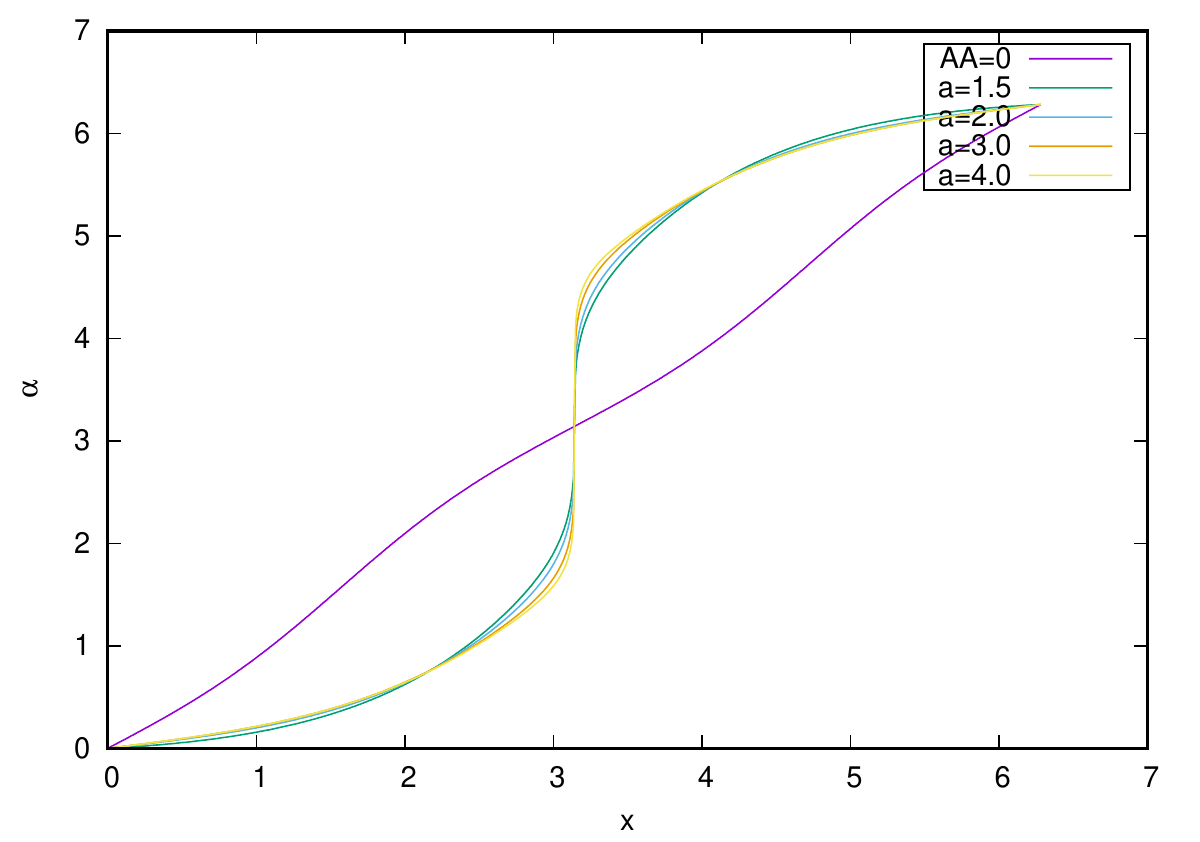}
         \caption{$n=2$ and $q=1$.}
         \label{fig:a}
     \end{subfigure}
     \hfill
     \begin{subfigure}[b]{0.45\textwidth}
         \centering
         \includegraphics[width=\textwidth]{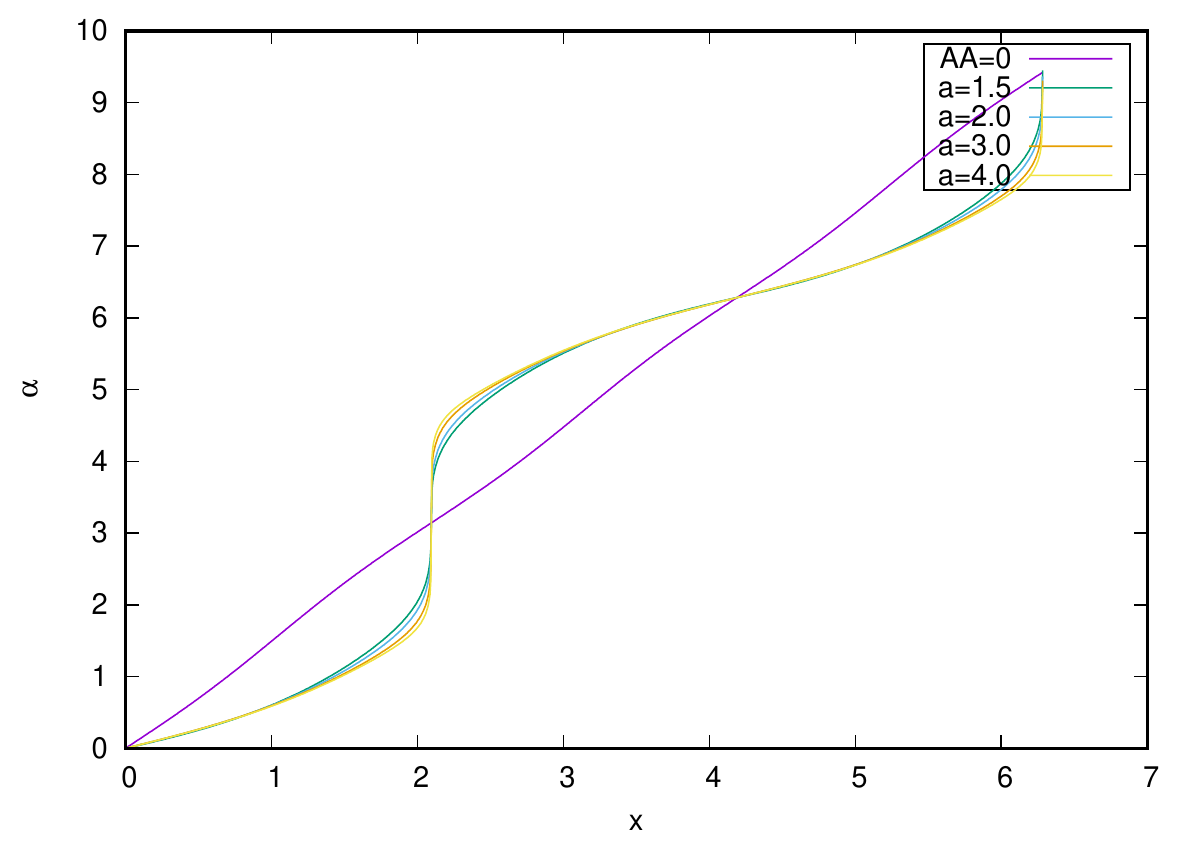}
         \caption{$n=3$ and $q=1$.}
         \label{fig:b}
     \end{subfigure}

     \vspace{0.5cm} % Espacio vertical opcional entre filas

     % FILA 2: Imágenes C y D
     \begin{subfigure}[b]{0.45\textwidth}
         \centering
         \includegraphics[width=\textwidth]{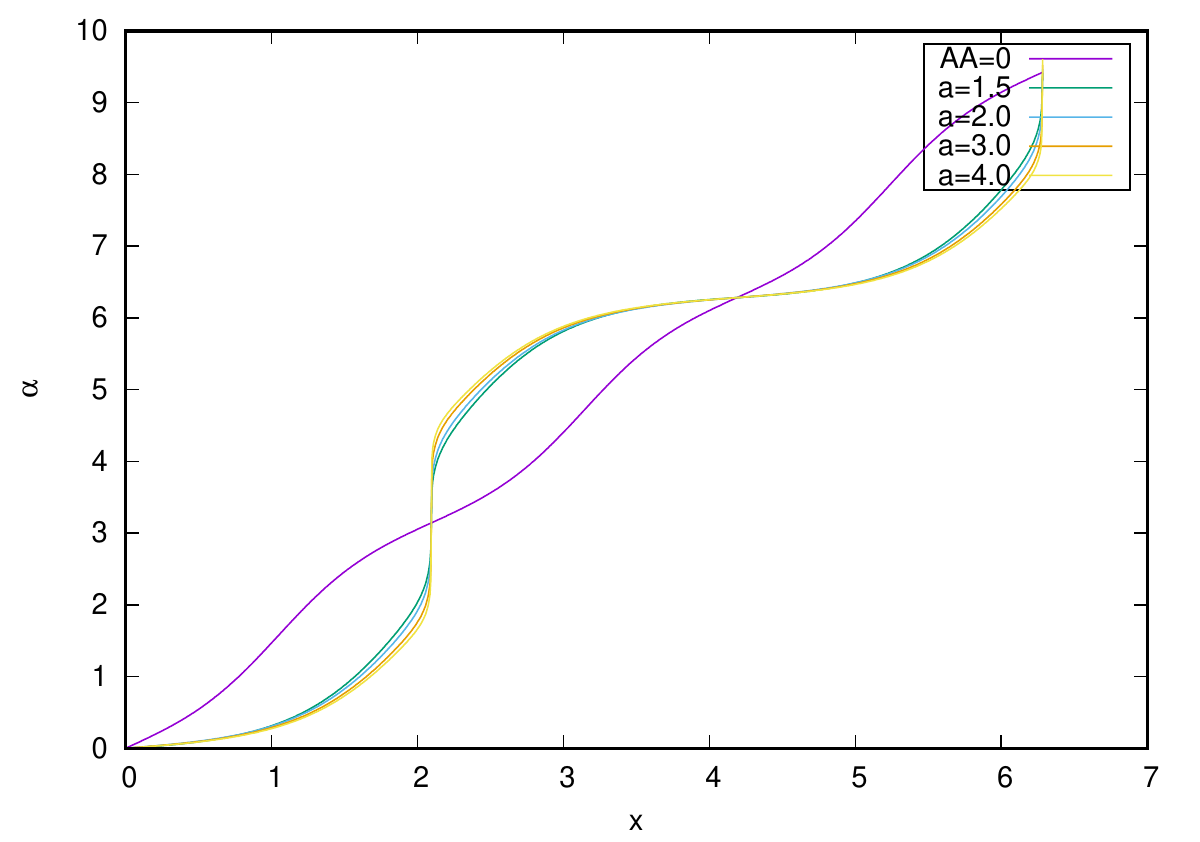}
         \caption{$n=3$ and $q=2$.}
         \label{fig:c}
     \end{subfigure}
     \hfill
     \begin{subfigure}[b]{0.45\textwidth}
         \centering
         \includegraphics[width=\textwidth]{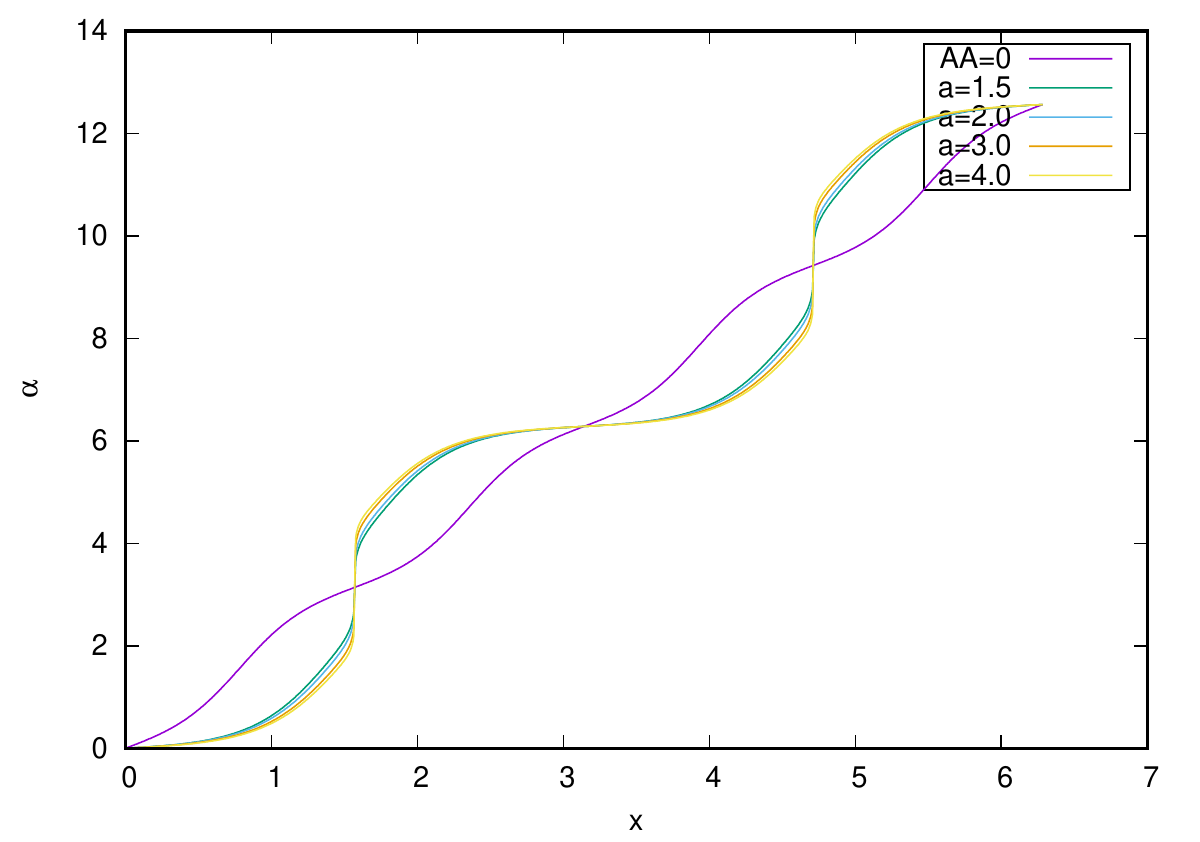}
         \caption{$n=4$ and $q=3$.}
         \label{fig:d}
     \end{subfigure}

     \caption{Solutions with potential $A_{0}\,\frac{\sin(\frac{\alpha}{2})^{r-1}}{\cos(\frac{\alpha}{2})^{r+1}}$.}
     \label{fig_old_solutions_extra}
\end{figure}

\begin{figure}
    \centering
    \captionsetup{type=figure}
     % FILA 1: Imágenes A y B
     \begin{subfigure}[b]{0.45\textwidth}
         \centering
         \includegraphics[width=\textwidth]{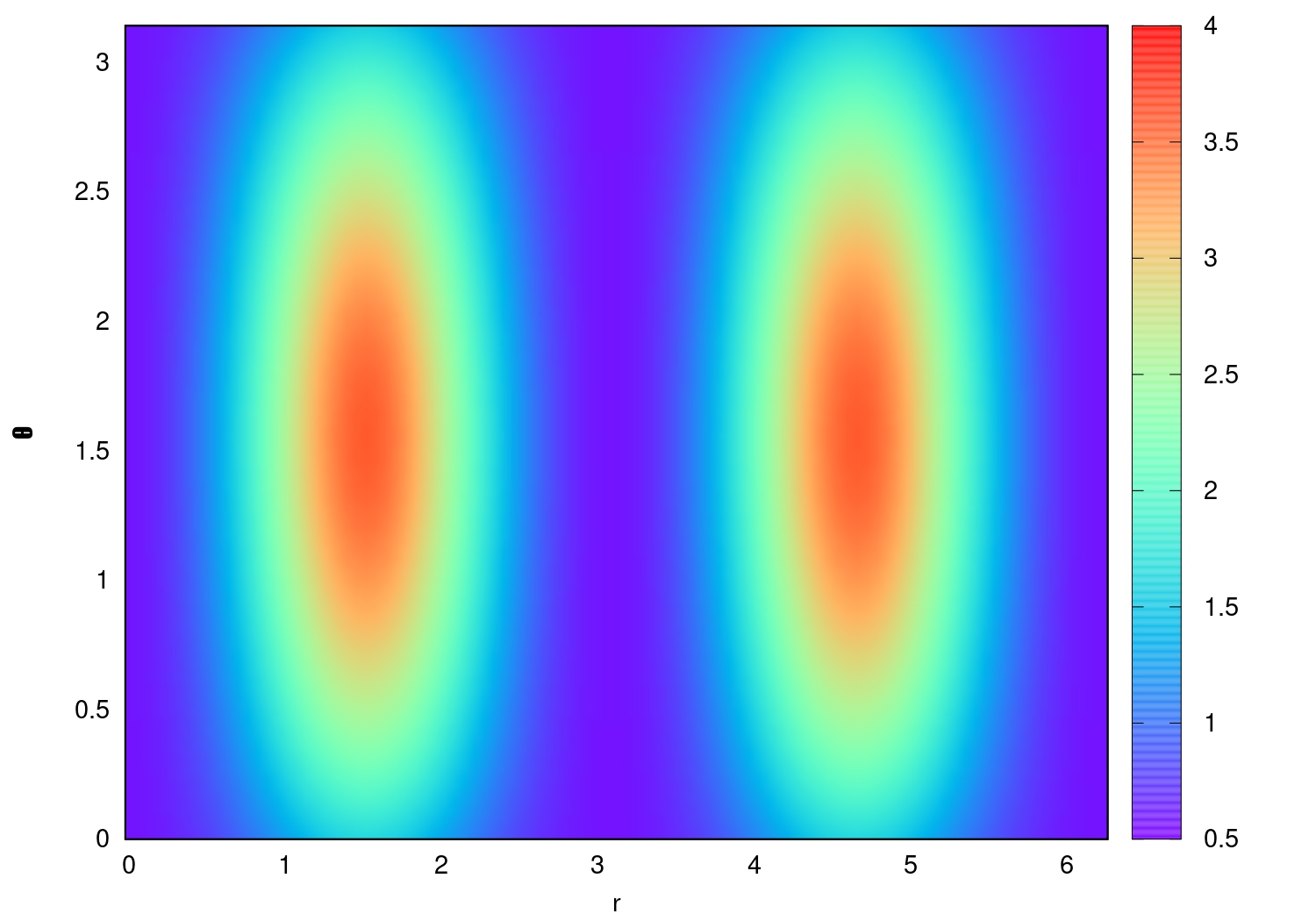}
         \caption{$A_{0}=0$.}
         \label{fig:e}
     \end{subfigure}
     \hfill
     \begin{subfigure}[b]{0.45\textwidth}
         \centering
         \includegraphics[width=\textwidth]{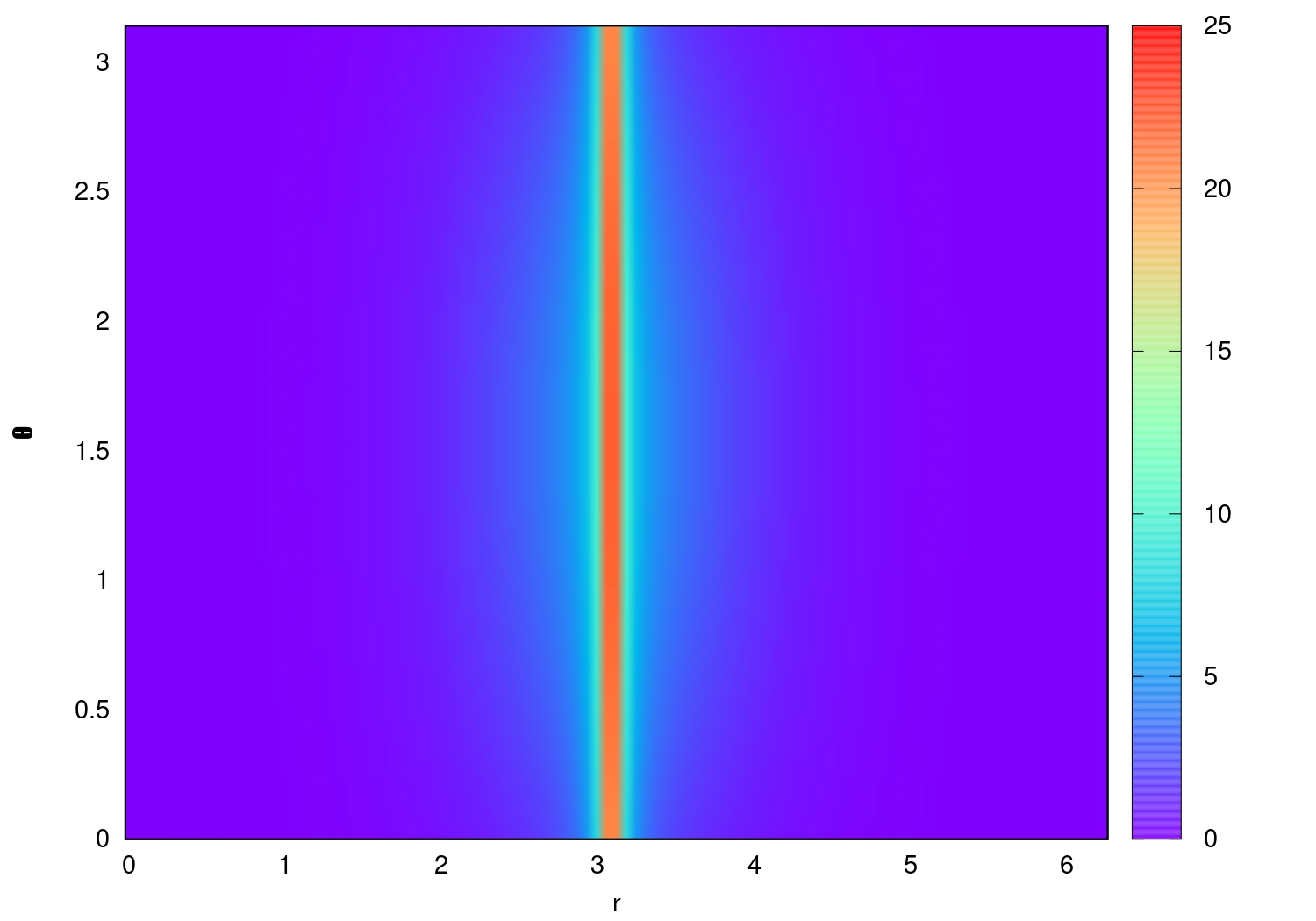}
         \caption{$r=1.5$.}
         \label{fig:f}
     \end{subfigure}

     \vspace{0.5cm} % Espacio vertical opcional entre filas

     % FILA 2: Imágenes C y D
     \begin{subfigure}[b]{0.45\textwidth}
         \centering
         \includegraphics[width=\textwidth]{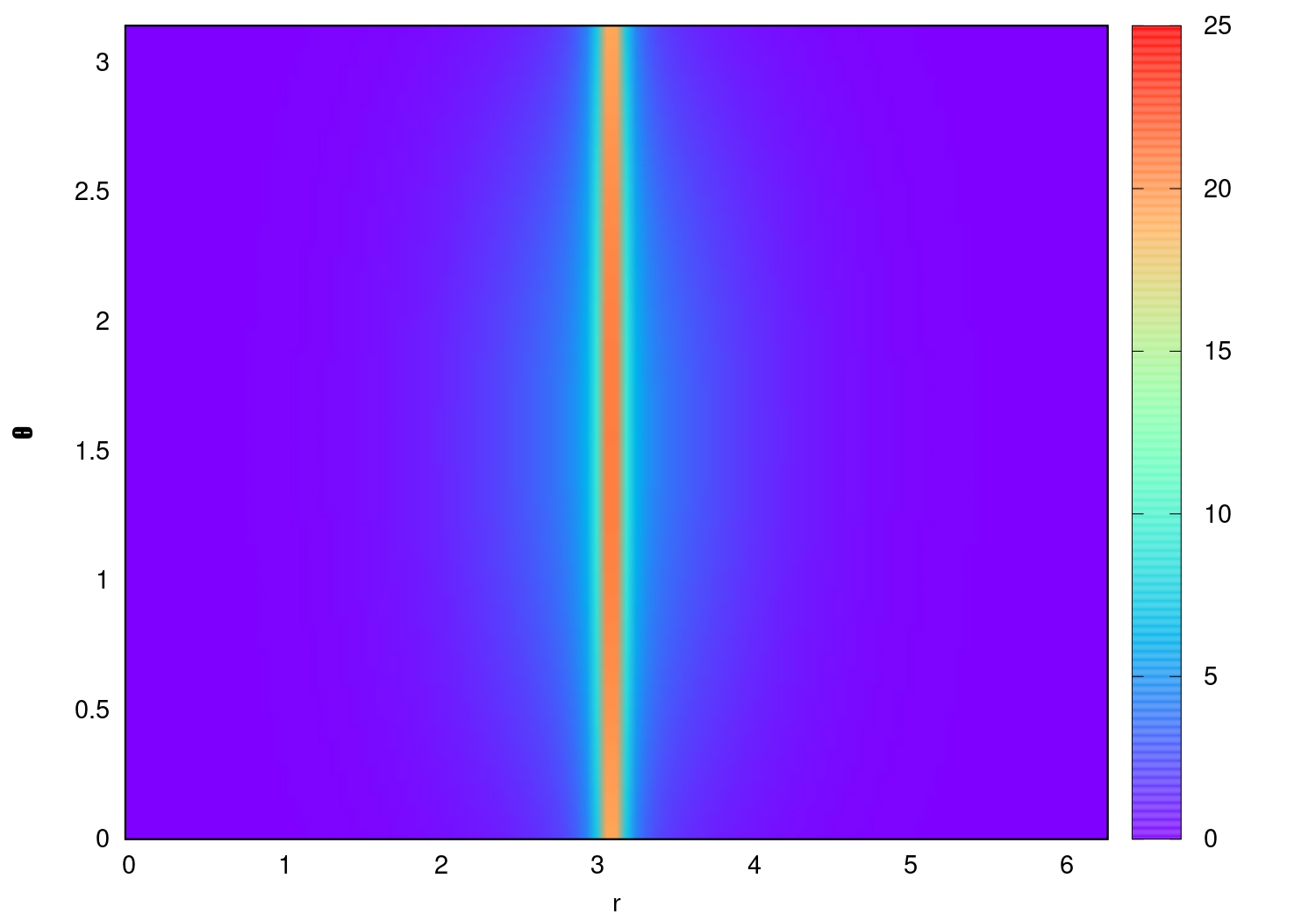}
         \caption{$r=2.0$.}
         \label{fig:g}
     \end{subfigure}
     \hfill
     \begin{subfigure}[b]{0.45\textwidth}
         \centering
         \includegraphics[width=\textwidth]{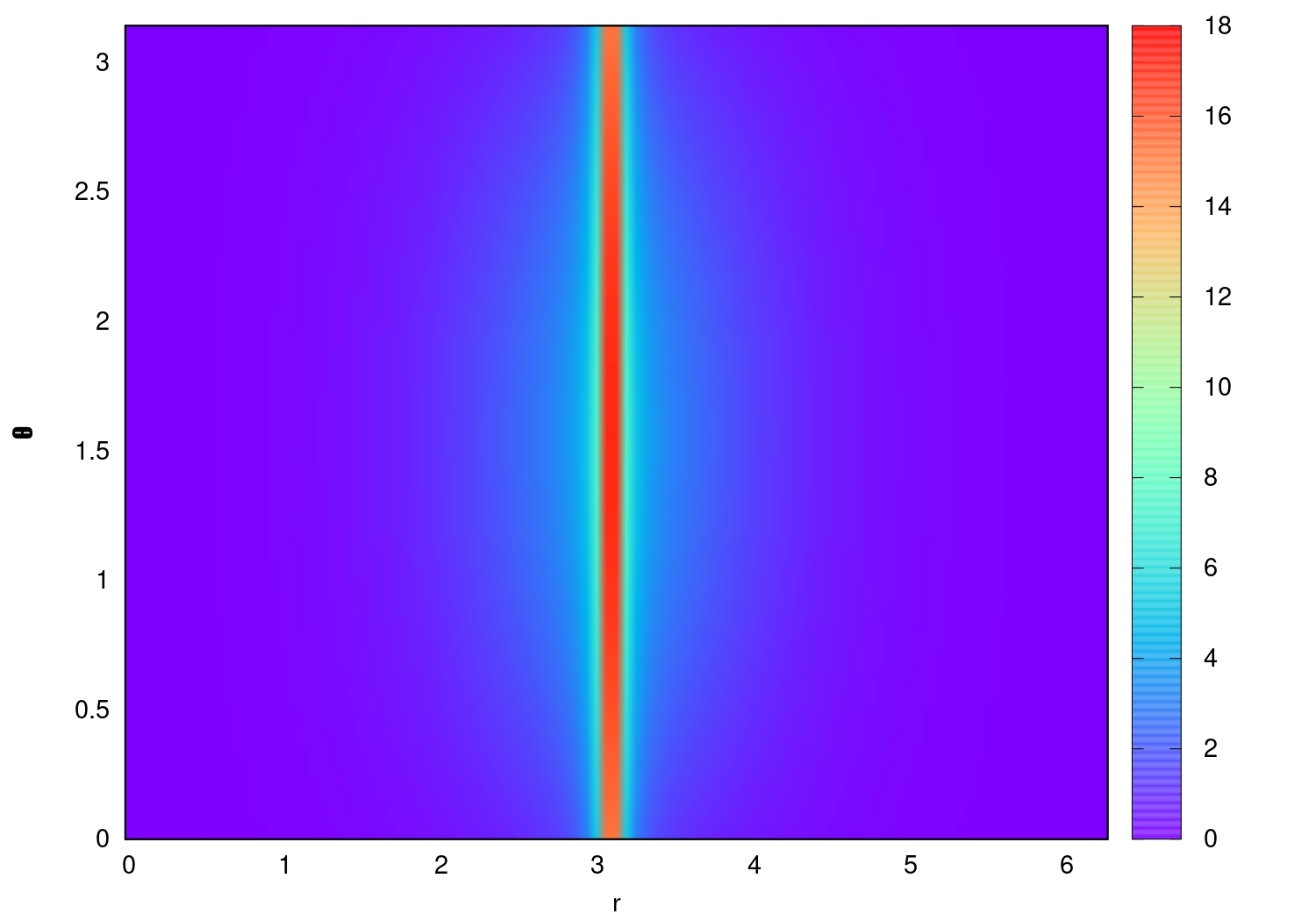}
         \caption{$r=3.0$.}
         \label{fig:h}
     \end{subfigure}
     \caption{Energy of the solutions with potential $A_{0}\,\frac{\sin(\frac{\alpha}{2})^{r-1}}{\cos(\frac{\alpha}{2})^{r+1}}$.}
     \label{fig_old_energy_extra}
\end{figure}

Finally, defining the current one-form $J=J_{x}\,dy-J_{y}\,dx$ and combining it with $\omega=\cos\alpha\,d\Theta$, we arrive at the generalized BPS bound for the reduced functional in the presence of potential terms:
\begin{equation}
E \ge \left| \int_{\partial \Sigma} (\omega + J) \right| \;.
\end{equation}
It is instructive to compare our BPS configurations with the numerical solutions constructed in Ref.~\cite{OldSolutions} in the presence of a potential term of the form Eq.~(\ref{effective_potential}). These solutions are depicted in Fig.~\ref{fig_old_solutions_extra} for representative values of $q$ and $n$ (see Section 2 of Ref. \cite{OldSolutions} for conventions), while Fig.~\ref{fig_old_energy_extra} illustrates their corresponding energies. Remarkably, the field profiles undergo a sharp structural change as soon as $A \neq 0$, indicating a non-trivial discontinuity in the solution space with respect to the parameter $A$.

In conclusion, the systematic inclusion of potential terms analyzed in this appendix demonstrates the robustness of the proposed BPS framework. Although specific potential profiles—such as the one parameterized by $G(1, a, \alpha)$—deviate from simple linear mass terms, they reveal a rich non-linear structure and point toward sharp transitions in the solution space when the deformation parameter becomes non-zero. These generalized BPS bounds and their associated current corrections provide a powerful extension of the theory, opening new avenues for investigating how explicit symmetry breaking and mass-like perturbations affect the stability and physical properties of baryonic tubes.

\section{Numerical Algorithm for the Baryon Quantization Manifold}
\label{app:quantization_manifold_code}

This appendix provides the self-contained computational scheme used to construct the two-dimensional quantization manifold $\mathcal{M}$ in the parameter space $(C_0, B_n) \in \mathbb{C} \times \mathbb{R} \cong \mathbb{R}^3$, ensuring that the topological charge satisfies $B = -n p$ within controlled numerical tolerances.

\subsection{Analytical Formulation on the Boundary}
On the unit circle boundary $L = \{ t = e^{i\phi} : \phi \in [0, 2\pi) \}$, the Blaschke product satisfies $|B(t)| = 1$, and its boundary phase $\Theta_{\text{top}}(\phi) = \arg B(e^{i\phi})$ accumulates a total winding of $2\pi n$. The regular background field generated via the Riemann-Hilbert problem evaluates on $L$ to:
\begin{equation}
    \mathcal{Z}_{\text{reg}}(e^{i\phi}) = \frac{C_0 + i B_n e^{in\phi} - \bar{C}_0 e^{2in\phi}}{\prod_{j=1}^n (1 - \bar{\zeta}_j e^{i\phi})^2} = X_{\text{reg}}(\phi) + i \Theta_{\text{reg}}(\phi).
\end{equation}
The total boundary field is $Z(e^{i\phi}) = X(\phi) + i\Theta(\phi)$, with $X(\phi) = X_{\text{reg}}(\phi)$ (since $X_{\text{top}} = \ln|B| = 0$) and $\Theta(\phi) = \Theta_{\text{top}}(\phi) + \Theta_{\text{reg}}(\phi)$. The net baryon charge enclosed within the domain is evaluated through the boundary 1-form via Stokes' theorem:
\begin{equation}
    B(C_0, B_n) = -\frac{p}{2\pi} \int_{0}^{2\pi} g[X(\phi)] \sin(\Theta(\phi)) \left( \frac{d\Theta}{d\phi} \right) d\phi,
\end{equation}
where $g[X] = 2\arctan(e^X) + \operatorname{sech}(X)\tanh(X)$. The quantization manifold corresponds to the zero-level set:
\begin{equation}
    \mathcal{F}(C_0, B_n) \equiv B(C_0, B_n) - (-n p) = 0.
\end{equation}

\subsection{Algorithmic Implementation and Precision Control}
Parameterizing $C_0 = A_0+iB_0$, the root-finding algorithm maps the implicit surface $B_n^*(A_0, B_0)$ such that $\mathcal{F}(A_0, B_0, B_n^*) = 0$ over a regular grid in $(A_0, B_0)$.

\begin{algorithm}[H]
\caption{Computation of the Quantization Manifold $\mathcal{M}$}
\label{alg:manifold}
\begin{algorithmic}
\REQUIRE Peak locations $\{\zeta_j\}_{j=1}^n$ with $|\zeta_j| < 1$; target charge $B_{\text{target}} = -n p$.
\REQUIRE Domain grid: $A_0 \in [{A_0}^{\min}, {A_0}^{\max}]$ ($N_{A_0}$ steps), $B_0 \in [{B_0}^{\min}, {B_0}^{\max}]$ ($N_{B_0}$ steps).
\REQUIRE Discretization: $N_\phi = 4096$ boundary points; $\phi_k = 2\pi k / N_\phi$ for $k=0, \dots, N_\phi - 1$.
\REQUIRE Precision criteria: Root tolerance $\text{tol}_{\text{root}} = 10^{-10}$; Max iterations $K_{\max} = 100$.
\ENSURE Quantization surface $B_n^*(A_0, B_0)$ and convergence flag matrix.
\FORALL{$(A_0, B_0)$ on the parameter grid}
    \STATE Define the 1D objective function $f(B_n) \equiv B(A_0+iB_0, B_n) - B_{\text{target}}$.
    \STATE Find a sign-changing bracket $[B_{n,\text{left}}, B_{n,\text{right}}]$ such that $f(B_{n,\text{left}}) \cdot f(B_{n,\text{right}}) \le 0$.
    \IF{bracket found}
        \STATE Compute $ B_n^* = \operatorname{BrentQ}\left( f, \, B_{n,\text{left}}, \, B_{n,\text{right}}, \, \text{xtol} = 10^{-10}, \, \text{rtol} = 10^{-12} \right)$.
        \STATE Store $B_n^*(u, v)$ and evaluate residual error $\delta B = |f(B_n^*)|$.
    \ELSE
        \STATE Mark $(u, v)$ as non-convergent (outside the topological bubble).
    \ENDIF
\ENDFOR
\end{algorithmic}
\end{algorithm}

\subsection{Quadrature Convergence and Numerical Accuracy}
To guarantee that the replication of these results is free of discretization ambiguities:
\begin{enumerate}
    \item \textbf{Boundary Quadrature:} The periodic integral over $\phi \in [0, 2\pi)$ is performed using the trapezoidal rule (which exhibits exponential convergence for smooth periodic functions on closed intervals) with $N_\phi = 4096$ nodes. The quadrature truncation error satisfies:
    \begin{equation}
        \epsilon_{\text{quad}} \equiv \left| \int_0^{2\pi} \omega(\phi) d\phi - \sum_{k=0}^{N_\phi-1} \omega(\phi_k) \Delta\phi \right| < 10^{-11}.
    \end{equation}
    \item \textbf{Root-Finding Precision:} The scalar root $B_n^*(u, v)$ is solved to an absolute parameter tolerance of $|B_n^{(k+1)} - B_n^{(k)}| \le 10^{-10}$ and a functional residual bound:
    \begin{equation}
        |B(C_0, B_n^*) - (-n p)| < 5 \times 10^{-10},
    \end{equation}
    guaranteeing that the integer quantization condition is satisfied down to single-precision machine epsilon limits.
    \item \textbf{Derivatives of Phase:} The derivative $d\Theta/d\phi$ is computed analytically by differentiating the logarithm of the finite Blaschke product and $\mathcal{Z}_{\text{reg}}$ directly, avoiding any loss of precision associated with numerical finite differences on boundary phases.
\end{enumerate}
\end{appendices}

\bibliography{biblio.bib}
\bibliographystyle{utphys}

\end{document}